\documentclass[onecolumn]{IEEEtran}
\usepackage{soul}
\usepackage[mathscr]{eucal}
\usepackage[cmex10]{amsmath}
\usepackage{epsfig,epsf,psfrag}
\usepackage{amssymb,amsmath,amsthm,amsfonts,latexsym}
\usepackage{amsmath,graphicx,bm,xcolor,url,overpic}
\usepackage[caption=false]{subfig} 
\usepackage{fixltx2e}
\usepackage{array}
\usepackage{verbatim}
\usepackage{bm}
\usepackage{algorithmic}
\usepackage{algorithm}
\usepackage{verbatim}
\usepackage{textcomp}
\usepackage{mathrsfs}
\usepackage{epstopdf}

\newcommand{\openone}{\leavevmode\hbox{\small1\normalsize\kern-.33em1}}

\catcode`~=11 \def\UrlSpecials{\do\~{\kern -.15em\lower .7ex\hbox{~}\kern .04em}} \catcode`~=13 

\allowdisplaybreaks[4]

\newcommand{\nn}{\nonumber}

\newcommand{\calA}{\mathcal{A}}
\newcommand{\calB}{\mathcal{B}}

\newcommand{\calI}{\mathcal{I}}

\newcommand{\calL}{\mathcal{L}}
\newcommand{\calM}{\mathcal{M}}

\newcommand{\calP}{\mathcal{P}}

\newcommand{\calS}{\mathcal{S}}
\newcommand{\calT}{\mathcal{T}}

\newcommand{\calV}{\mathcal{V}}
\newcommand{\calW}{\mathcal{W}}
\newcommand{\calX}{\mathcal{X}}

\newcommand{\bP}{\mathbf{P}}

\newcommand{\bQ}{\mathbf{Q}}

\newcommand{\bX}{\mathbf{X}}

\newcommand{\rmH}{\mathrm{H}}

\newcommand{\rmr}{\mathrm{r}}

\newcommand{\bbN}{\mathbb{N}}

\newcommand{\bbP}{\mathbb{P}}

\newcommand{\bbR}{\mathbb{R}}

\DeclareMathAlphabet{\mathbsf}{OT1}{cmss}{bx}{n}
\DeclareMathAlphabet{\mathssf}{OT1}{cmss}{m}{sl}

\DeclareSymbolFont{bsfletters}{OT1}{cmss}{bx}{n}  
\DeclareSymbolFont{ssfletters}{OT1}{cmss}{m}{n}
\DeclareMathSymbol{\bsfGamma}{0}{bsfletters}{'000}
\DeclareMathSymbol{\ssfGamma}{0}{ssfletters}{'000}
\DeclareMathSymbol{\bsfDelta}{0}{bsfletters}{'001}
\DeclareMathSymbol{\ssfDelta}{0}{ssfletters}{'001}
\DeclareMathSymbol{\bsfTheta}{0}{bsfletters}{'002}
\DeclareMathSymbol{\ssfTheta}{0}{ssfletters}{'002}
\DeclareMathSymbol{\bsfLambda}{0}{bsfletters}{'003}
\DeclareMathSymbol{\ssfLambda}{0}{ssfletters}{'003}
\DeclareMathSymbol{\bsfXi}{0}{bsfletters}{'004}
\DeclareMathSymbol{\ssfXi}{0}{ssfletters}{'004}
\DeclareMathSymbol{\bsfPi}{0}{bsfletters}{'005}
\DeclareMathSymbol{\ssfPi}{0}{ssfletters}{'005}
\DeclareMathSymbol{\bsfSigma}{0}{bsfletters}{'006}
\DeclareMathSymbol{\ssfSigma}{0}{ssfletters}{'006}
\DeclareMathSymbol{\bsfUpsilon}{0}{bsfletters}{'007}
\DeclareMathSymbol{\ssfUpsilon}{0}{ssfletters}{'007}
\DeclareMathSymbol{\bsfPhi}{0}{bsfletters}{'010}
\DeclareMathSymbol{\ssfPhi}{0}{ssfletters}{'010}
\DeclareMathSymbol{\bsfPsi}{0}{bsfletters}{'011}
\DeclareMathSymbol{\ssfPsi}{0}{ssfletters}{'011}
\DeclareMathSymbol{\bsfOmega}{0}{bsfletters}{'012}
\DeclareMathSymbol{\ssfOmega}{0}{ssfletters}{'012}

\newcommand{\tilE}{\tilde{E}}

\newcommand{\hatG}{\hat{G}}

\newcommand{\tilG}{\tilde{G}}

\newcommand{\tilP}{\tilde{P}}

\newcommand{\hatT}{\hat{T}}

\newcommand{\bari}{\bar{i}}

\newtheorem{theorem}{Theorem} 
\newtheorem{lemma}[theorem]{Lemma}

\usepackage{graphicx,cite}
\usepackage{epstopdf}
\usepackage{enumerate}
\usepackage{color}
\usepackage{bbm}
\usepackage{booktabs}
\usepackage{makecell}
\usepackage[ colorlinks = true,
             linkcolor = blue,
             urlcolor  = blue,
             citecolor = red,
             anchorcolor = green,]{hyperref}

\newcommand{\blue}[1]{\textcolor{blue}{#1}}

\makeatletter

\newcommand{\Rmnum}[1]{\expandafter\@slowromancap\romannumeral #1@}
\makeatother

\usepackage{setspace}
\def\BibTeX{{\rm B\kern-.05em{\sc i\kern-.025em b}\kern-.08em
    T\kern-.1667em\lower.7ex\hbox{E}\kern-.125emX}}
\def\BibTeX{{\rm B\kern-.05em{\sc i\kern-.025em b}\kern-.08em T\kern-.1667em\lower.7ex\hbox{E}\kern-.125emX}}
\begin{document}

\title{Achievable Error Exponents for Two-Phase Multiple Classification}

\author{Lin Zhou, Jun Diao and  Lin Bai

\thanks{This work was partially presented at IEEE ISIT 2022~\cite{bai2022achievable}, IEEE ICASSP 2023~\cite{diao2023achievable} and IEEE ISIT 2023~\cite{diao2023classification}.}

\thanks{The authors are with the School of Cyber Science and Technology, Beihang University, Beijing, China, 100083 (Emails: \{lzhou, jundiao, l.bai\}@buaa.edu.cn). L. Zhou and L. Bai are also with the Beijing Laboratory for General Aviation Technology, Beihang University, Beijing.}

\thanks{This work was supported in part by the National Natural Science Foundation of China under Grants 61922010 and 62201022,  in part by the Beijing Natural Science Foundation under Grant JQ20019 and 4232007 and in part by funds of Beihang university.}
}

\maketitle

\begin{abstract}
\blue{
We revisit $M$-ary classification of Gutman (TIT 1989), where one is tasked to determine whether a testing sequence is generated with the same distribution as one of the $M$ training sequences or not. Our main result is a two-phase test, its theoretical analysis and its optimality guarantee. Specifically, our two-phase test is a special case of a sequential test with only two decision time points: the first phase of our test is a fixed-length test with a reject option, the second-phase of our test proceeds only if a reject option is decided in the first phase and the second phase of our test does \emph{not} allow a reject option. To provide theoretical guarantee for our test, we derive achievable error exponents using the method of types and derive a converse result for the optimal sequential test using the techniques recently proposed by Hsu, Li and Wang (ITW, 2022) for binary classification. Analytically and numerically, we show that our two phase test achieves the performance of an optimal sequential test with proper choice of test parameters. In particular, similarly as the optimal sequential test, our test does not need a final reject option to achieve the optimal error exponent region while an optimal fixed-length test needs a reject option to achieve the same region. Finally, we specialize our results to binary classification when $M=2$ and to $M$-ary hypothesis testing when the ratio of the lengths of training sequences
and testing sequences tends to infinity so that generating distributions can be estimated perfectly. For both specializations, our two-phase test bridges over optimal fixed-length and sequential tests in the same spirit as the results of Lalitha and Javidi for binary hypothesis testing (ISIT 2016). In summary, our results generalize the design and analysis of the two-phase test for binary hypothesis testing to broader and more practical families of statistical classification with multiple decision outcomes.
}
\end{abstract}


\section{Introduction}

Binary hypothesis testing lies in the intersection of information theory and statistics with applications in various domains~\cite{csiszar2004it}. In this problem, one is given two distributions $P_1$ and $P_2$ and a test sequence $Y^n$. The task is to decide from which distribution the test sequence $Y^n$ is generated i.i.d. from and the performance critera are the type-I and type-II error probabilities. The Neyman-Pearson lemma~\cite{cover2012elements} states that the likelihood ratio test (LRT) is optimal. When the type-I error probability is upper bounded by a constant, the Chernoff-Stein lemma~\cite{chernoff1952measure} shows that the type-II error probability decays exponentially fast at the speed of $D(P_1\|P_2)$. Blahut~\cite{blahut1974hypothesis} showed that to ensure that the type-I error probability decays exponentially fast with a speed $\lambda$, the decay rate of the type-II error probability is reduced from $D(P_1\|P_2)$ to $\min_{\tilP: D(\tilP\|P_1)\leq \lambda}D(\tilP\|P_2)$. Such a tradeoff can be improved in the sequential setting where the length of the test sequence is allowed to vary but the expected length is upper bounded. In this setting, Wald~\cite{wald1948optimum} showed that the sequential probability ratio test (SPRT) is optimal and achieves maximal error exponents for both types of error probabilities \emph{simultaneously}.

The superior performance of the SPRT comes at the high complexity of the test design, where at each time point, one takes a new test sample and determines whether to continue collecting samples or to make a decision. \blue{One might wonder whether it is possible to achieve the performance of the SPRT while simplifying the test design with limited number of decision time points.} Lalitha and Javidi~\cite{lalithaalmost} answered this question affirmatively by proposing a two-phase test. Specifically, the test in~\cite{lalithaalmost} consists of two phases: the first phase is a fixed-length test taking $n$ samples, the second phase is another fixed-length test taking additional $\lceil (k-1)n\rceil$\footnote{In the rest of the paper, for simplicity, we drop the integer constraint and use $(k-1)n$ and $kn$.} samples for some real number $k\geq 1$ and the second phase proceeds only if the first phase outputs a reject option, indicating that $n$ samples are not sufficient to make a reliable decision. With proper design, the second phase proceeds with an exponentially small probability. \blue{Thus, the two-phase test is simply a sequential test with two decision time points $n$ and $kn$ and the asymptotical average sample size is $n$. Such a design renders the two-phase test more practical than a usual sequential test that checks at each time point whether to stop after collecting each data sample. It is the ease of the test design that motivates the study of the two-phase test for hypothesis testing, analogously to the motivation for the two-phase code of Yamamoto and Itoh~\cite{yamamoto1979asymptotic} for channel coding with feedback.}

\blue{
The theoretical results for fixed-length and sequential tests for binary hypothesis testing have been generalized to the $M$-ary hypothesis testing~\cite{leang1997asymptotics,tuncel2005error,baum1994sequential}. For the more practical problem where the generating distributions are \emph{unknown}, Gutman~\cite{gutman1989asymptotically} proposed the statistical classification framework where one determines whether a test sequence $Y^n$ is generated from one of the unknown distributions using empirically observed statistics. Gutman proposed a distribution free test, derived its achievable error exponent under each hypothesis and proved its optimality under the generalized Neyman-Pearson criterion. Recently, Haghifam, Tan and Khisti~\cite{mahdi2021sequential} generalized Gutman's result to the sequential setting and derived the achievable error exponents under the universal error probability constraint. The results for the binary case of \cite{mahdi2021sequential} was further generalized by Hsu, Li and Wang~\cite{Ihwang2022sequential}, who derived tight results with matching achievability and converse bounds for both the expected stopping time universality constraint and error probability universality constraint.
}

However, there is no publication that derives the performance of a two-phase test for the more practical problem of statistical classification. Inspired by the results of Lalitha and Javidi~\cite{lalithaalmost} for binary hypothesis testing, we are interested in the following question. Can one propose a two-phase test for statistical classification, and demonstrate that the test has performance close to the optimal sequential test using the simple design of a fixed-length test? Fortunately, we answer the question affirmatively. Our contributions are summarized as follows.

\subsection{Main Contributions}
\blue{
We generalize the results of Lalitha and Javidi~\cite{lalithaalmost} for binary hypothesis testing to the more practical statistical classification problem where the generating distributions under each hypothesis are \emph{unknown} and there are more than two hypotheses. We propose a two-phase tests using the generalized Jensen-Shannon divergence~\cite[Eq. (2.3)]{zhou2020second} that takes empirical distributions of training sequences and the testing sequence as the parameters. We derive the achievable error exponents of our tests using the method of types. To provide the optimality guarantee of the two-phase test, we characterize the error exponent of an optimal sequential test under error probability universality constraint using the techniques recently proposed by Hsu, Li and Wang for binary sequential classification~\cite{Ihwang2022sequential}. This is because our two-phase test is a special case of the sequential test and thus the converse bound for sequential classification is naturally a converse bound for our two-phase test. Analytical results and numerical simulations verify that our two-phase tests achieve error exponent close to the optimal sequential test and bridge the performance gaps between the fixed-length test~\cite{gutman1989asymptotically} and the sequential test. In particular, our two-phase test does not require a reject option to achieve the optimal error exponent region of the sequential test, which is in stark contrast with the fixed-length test~\cite{gutman1989asymptotically}. The non-asymptotic complexity of our test is at most multiple of a fixed-length test and the asymptotic complexity of our test is the same as the fixed-length test. Furthermore, we specialize our results to binary classification with $M=2$ and compare the performance between the specialized test and the threshold-based test in our previous conference paper~\cite[Eq. (14) and Eq. (15)]{bai2022achievable}. Numerical results demonstrate that the specialized test in this paper is superior to the threshold-based test in the Bayesian setting.
}

\blue{
We further discuss how the performance of our tests are influenced by $\alpha$, the ratio of the lengths of training sequences and that of the testing sequence. We demonstrate that the more the training sequences (larger $\alpha$), the better the performance. When $\alpha\to\infty$ so that the generating distribution under each hypothesis is estimated accurately, our results specialize to $M$-ary hypothesis testing. Analytical results and numerical simulations show that by tuning the design parameters, the achievable error exponents of our test approach either the fixed-length test~\cite{tuncel2005error} or the sequential test~\cite{baum1994sequential} in both the Neyman-Pearson and the Bayesian settings. Note that our test is different from the LRT-based test of Lalitha and Javidi~\cite[Eq. (14) and Eq. (15)]{lalithaalmost} when specialized to $M=2$, but it is more amenable to generalizations to the case with unknown distributions and multiple hypotheses.
}

\subsection{Other Related Works}
\blue{
We remark that the two-phase test for hypothesis testing is closely related with the two-phase code for channel coding with feedback~\cite{schalkwijk1966coding,burnashev1976data} by Yamamoto and Itoh~\cite{yamamoto1979asymptotic}. Specifically, the two-phase code in~\cite{yamamoto1979asymptotic} includes a message phase and a confirmation phase. In the message phase, a message is transmitted using a fixed number $n$ of channel uses. After receiving the channel outputs, the receiver checks whether the message is transmitted correctly or not and sends either an ACK or NACK signal to inform the transmitter. Once the NACK signal is received, the transmitter retransmits the message. Note that the ACK or NACK signal corresponds to the non-reject and reject decisions of the first phase for hypothesis testing. However, we would like to note that the second phase of both problems differ. In channel coding, the second phase is a retransmission of the same message. In contrast, in hypothesis testing, the second phase collects another $(k-1)n$ data samples and makes a final decision using all $kn$ data samples from both phases. Furthermore, since the objectives of both problems are different, the analyses are rather different and thus the theoretical analyses for channel coding with feedback~\cite{burnashev1976data,naghshvar2012optimal,yang2022sequential} does not directly imply the results for hypothesis testing, regardless whether the generating distribution is known or not.
}

We then provide a non-exhaustive list of related works on statistical classification. Zhou, Tan and Motani~\cite{zhou2020second} analyzed the finite sample performance of Gutman's test. Hsu and Wang~\cite{hsu2020binary} generalized Gutman's results to the mismatched setting where the generating distributions of the training sequences and the test sequences deviate slightly.
The Bayesian error probability of binary classification was studied by Merhav and Ziv~\cite{merhav1991bayesian} in the asymptotic case and more recently by Saito~\cite{shota2020beyasian,shota2021beyasian} in the finite blocklength using Bayes codes. Gutman's results are also generalized to multiple test sequences~\cite{unnikrishnan2015asymptotically}, distributed detection~\cite{he2019distributed}, quickest change-point detection~\cite{he2021optimal}, outlier hypothesis testing~\cite{li2014,li2017universal} and universal sequential classification~\cite{Ihwang2022sequential}.

\subsection*{Notation}
We use $\bbR$, $\bbR_+$, $\bbN$ to denote the set of real numbers, non-negative real numbers, and natural numbers respectively. Given any integer $a\in\bbN$, we use $[a]$ to denote $[1,2,\ldots,a]$. Random variables and their realizations are denoted by upper case variables (e.g., $X$) and lower case variables (e.g., $x$), respectively. All sets are denoted in calligraphic font (e.g., $\mathcal{X}$). For any $N\in\bbN$, let $X^N:=(X_1,\ldots X_N)$ be a random vector of length $N$ and let $x^N=(x_1,\ldots,x_N)$ be a particular realization of $X^N$. The set of all probability distributions on a finite set $\calX$ is denoted as $\calP(\calX)$.  We use $\mathbb{E}[\cdot]$ to denote expectation. Given a sequence $x^n\in\calX^n$, the type or empirical distribution is defined as $\hatT_{x^n}(a)=\frac{1}{n}\sum_{i=1}^{n}\mathbbm{1}(x_i)=a,\forall a\in\calX$. The set of types formed from length-$n$ sequences with alphabet $\calX$ is denoted as $\calP^n(\calX)$. Given any $P\in\calP^n(\calX)$, the set of all sequences of length $n$ with type $P$, the type class, is denoted as $\calT_P^n$.

\section{Problem Formulation, Fixed-length Test and Sequential Test}

\subsection{Problem Formulation}
In $M$-ary classification, we are given $M$ training sequences $\bX^N:=(X_1^N,\ldots,X_M^N)\in(\calX^N)^M$ of length $N\in\bbN$ generated i.i.d. according to distinct unknown distributions $\bP=(P_1,\ldots,P_M)\in\calP(\calX)^M$ and a test sequence $Y^\tau$ generated i.i.d. according to one of the $M$ distributions, where $\tau$ is a random stopping time with respect to the filtration $\sigma\{\bX^N,Y_1,\ldots Y_n\}$. For simplicity, we assume that $N=\lceil n\alpha\rceil$\footnote{In subsequent analysis, without loss of generality, we ignore the integer requirement and drop the ceiling operation for simplicity.}. The task is to design a test $\Phi=(\tau,\phi_\tau)$, which consists of a random stopping time $\tau$ and a mapping $\phi_\tau$: $\calX^{MN}\times\calX^\tau\rightarrow\{\rmH_1,\rmH_2,\ldots,\rmH_M\}$, to decide among the following $M$ hypotheses:
\begin{itemize}
\item $\rmH_j$: the test sequence $Y^\tau$ is generated i.i.d. from the same distribution as the $j^{\rm{th}}$ training sequence $X_j^N$,~$j\in[M]$.
\end{itemize}
At the random stopping time $\tau$, any mapping $\phi_\tau$ partitions the sample space into $M$ disjoint regions: $\{\calA_j(\phi_\tau)\}_{j\in[M]}$ where $Y^\tau \in \calA_j(\phi_\tau)$ favors hypothesis $\rmH_j$.

Given any test $\Phi$ and any tuple of distributions $\bP\in\calP(\calX)^M$, we have the following $M$ error probabilities to evaluate the performance of the test:
\begin{align}
\beta_j(\Phi|\bP):=\bbP_j \big\{\phi_\tau(\bX^N,Y^\tau)\neq\rmH_j\big\},~j\in[M],
\end{align}
where for each $j\in[M]$, we define $\bbP_j\{\cdot\}:=\operatorname{Pr}\{\cdot|\rmH_j\}$ under which $X_i^N\sim P_i$ for all $i\in[M]$ and $Y^\tau\sim P_j$. It is challenging to characterize the non-asymptotic performance of error probabilities with finite sample size. As a compromise, to be consistent with literature, one usually derives error exponents, i.e., the decay rates of error probabilities. For $M$-ary classification, we are interested the following error exponents
\begin{align}\label{Maryexponent}
E_j(\Phi|\bP):=\liminf_{n\to\infty}\frac{-\log\beta_j(\Phi|\bP)}{\mathbb{E}_j[\tau]},
\end{align}
and Bayesian error exponent:
\begin{align}\label{Beyasianexponent}
E_\mathrm{Bayesian}\big(\Phi|\bP\big):=\liminf_{n\to\infty}\frac{-\log \sum_{j=1}^{M}\pi_j\beta_j\big(\Phi|\bP\big)}{\mathbb{E}_j[\tau]},
\end{align}
where $\pi_j\in(0,1)$ is the prior probability of hypothesis $\rmH_j$ for each $j\in[M]$ such that $\sum_{j=1}^{M}\pi_j=1$.
Note that in both \eqref{Maryexponent} and \eqref{Beyasianexponent}, the random stopping time $\tau$ is a function of $n$ through the filtration $\sigma\{Y_1,\ldots,Y_n\}$ and thus the definitions are valid.

Fix any integer $n\in\bbN$. When the random stopping time $\tau$ is fixed as a constant such that $\tau=n$, the test is called a fixed-length test; when the random stopping time $\tau$ satisfies $\limsup_{n\to\infty}\frac{\mathbb{E}[\tau]}{n}\leq 1$, the test is called a sequential test. The achievable error exponents of a fixed-length test was derived by Gutman~\cite{gutman1989asymptotically}. We generalize the achievability and converse results of the sequential test for binary classification~\cite{Ihwang2022sequential} to $M$-ary classification.

\subsection{Fixed-length Test}
We first recall the results of the fixed-length test by Gutman~\cite{gutman1989asymptotically}.To present Gutman's test, we need the following measure of distributions. Given any two distributions $(P,Q)\in\calP(\calX)^2$ and any positive real number $\alpha\in\bbR_+$, the generalized Jensen-Shannon divergence is defined as
\begin{align} \label{GJS}
{\rm GJS}(P,Q,\alpha)&\!=\!\alpha D\left(\!P\Big\| \frac{\alpha P+Q}{1+\alpha}\!\right)\!+\!D\left(\!Q\Big \| \frac{\alpha P+Q}{1+\alpha}\!\right).
\end{align}
Then define the set
\begin{align}\label{Mdis}
\calM_{\mathrm{dis}}:=\{(i,j)\in[M]^2:i\neq j\}.
\end{align}
For any positive real number $\lambda\in\bbR_+$, Gutman's test $\Phi_{\mathrm{Gut}}^{(M)}=\big(n,\phi_{n,\mathrm{Gut}}^{(M)}\big)$ has a fixed stopping time $\tau=n$ and proceeds as follows using the training and test sequences $(\bX^N,Y^n)$:
\begin{align}\label{Gutman}
\phi_{n,\mathrm{Gut}}^{(M)}(\bX^N,Y^n)=\left\{\begin{array}{l}
\mathrm{H_1}\;\;\mathrm{if}~\mathrm{GJS}\big(\hatT_{X_i^N},\hatT_{Y^n},\alpha\big)\ge\lambda,\forall~i\in[2,M],\\*
\mathrm{H}_i\;\;\mathrm{if}~\mathrm{GJS}\big(\hatT_{X_i^N},\hatT_{Y^n},\alpha\big)<\lambda,\exists~i\in[1,M]\\*
\qquad\;\;\;\mathrm{and}~\mathrm{GJS}\big(\hatT_{X_t^N},\hatT_{Y^n},\alpha\big)\ge\lambda,\forall~t\in[M]\backslash i,\\*
\mathrm{H_r}\;\;\mathrm{otherwise},
\end{array}\right.
\end{align}
where $\rmH_\rmr$ denotes the reject option, which indicates that a reliable decision could not be made.

Gutman proved the following result.
\begin{theorem}\label{gut:mary}
For any tuple of distributions $\bP\in\calP(\calX)^M$ and any $\lambda\in\bbR^+$, Gutman's test satisfies
\begin{align}
E_j\big(\Phi_{\mathrm{Gut}}^{(M)}|\bP\big)\ge\lambda,j\in[M].
\end{align}
Furthermore, among all fixed-length tests that achieve the same exponent as Gutman's test, Gutman's test has the smallest rejection region.
\end{theorem}
Note that the largest achievable exponent $\lambda\in\bbR^+$ of Gutman's test such that the rejection probability vanishes as $n\to\infty$ is $\bar\lambda=\min\limits_{(i,j)\in\calM_{\mathrm{dis}}}\mathrm{GJS}(P_i,P_j,\alpha)$.

\subsection{Sequential Test}
\blue{
To present the result for sequential classification, we need the following definitions. Given $\beta\in(0,1)$ and any sequential test $\Phi$, we say that
$\Phi$ satisfies the universality constraint on the error probability with $\beta$ if for all tuple of distributions $\tilde\bP\in\calP^M(\calX)$,
\begin{align}\label{sequential_universal}
\max_{j\in[M]}\beta_j(\Phi|\tilde\bP)\le\beta.
\end{align}
For each $j\in[M]$, the type-$j$ error exponent is defined as
\begin{align}
E_j(\Phi^{(M)}|\tilde\bP):=&\liminf_{\beta\to 0}\frac{-\log\beta}{\mathbb{E}_j[\tau]}.
\end{align}
We propose a sequential test $\Phi_{\rm{seq}}^{(M)}$ satisfying the error probability universality for $M$-ary classification by generalizing the test in \cite{hsu2020binary}.
Given $\beta\in(0,1)$ and $n\in\bbN$, define the set
\begin{align}
\Psi_n:=\Big\{i\in[M]:~\mathrm{GJS}\big(\hatT_{X_i^N},\hatT_{Y^n},\alpha\big)>g(\beta,n)\Big\},
\end{align}
where
\begin{align}\label{gbeta}
g(\beta,n):=\frac{-\log\big(\beta(|\calX|-1)\big)}{n}+\frac{2|\calX|\log(n+1)}{n}+\frac{|\calX|\log(n\alpha+1)}{n}.
\end{align}
The sequential test $\Phi_{\rm{seq}}^{(M)}=\big(\tau_{\mathrm{seq}}^{(M)},\phi_{\tau_\mathrm{seq}}^{(M)}\big)$ consists of the random stopping time
\begin{align}
\tau_{\mathrm{seq}}:=\inf\big\{n\in\bbN:~|\Psi_n|\ge M-1\big\},
\end{align}
and the decision rule
\begin{align}\label{seqtest}
\phi_{\tau_\mathrm{seq}}^{(M)}(\bX^{N_{\tau_{\mathrm{seq}}}},Y^{\tau_{\mathrm{seq}}})=\mathrm{H}_j,~j=[M]\backslash\Psi_{\tau_{\mathrm{seq}}}.
\end{align}
We remark that under hypothesis $\rmH_j$, the scoring function $\mathrm{GJS}(\hatT_{X_j^N},\hatT_{Y^n},\alpha)$ tends to zero while all other $M-1$ scoring functions $\mathrm{GJS}(\hatT_{X_i^N},\hatT_{Y^n},\alpha),i\in\calM_j$ tend to the positive number $\mathrm{GJS}(P_i,P_j,\alpha)$ as the length of the testing sequence $n$ increases. Thus the size of $\Psi_{\tau_{\mathrm{seq}}}$ will not exceed $M-1$ asymptotically.
}

\blue{
We derive the achievable error exponent and prove its optimality in the similar spirit of~\cite{Ihwang2022sequential} as follows.
\begin{theorem}\label{wang:mary}
For any tuple of distributions $\bP\in\calP(\calX)^M$ and each $j\in[M]$, the above sequential test satisfying the error probability universality constraint achieves the error exponent as follows:
\begin{align}
E_j\big(\Phi_{\mathrm{seq}}^{(M)}|\bP\big)\ge\min\limits_{i\in\calM_j}\mathrm{GJS}(P_i,P_j,\alpha),
\end{align}
where $\calM_j$ is defined as the set $\{i\in[M]:i\neq j\}$. Conversely, for any family of tests $\Phi_n^{(M)}$ satisfying the error probability universality constraint,
\begin{align}
E_j\big(\Phi_n^{(M)}|\bP\big)\le\min\limits_{i\in\calM_j}\mathrm{GJS}(P_i,P_j,\alpha).
\end{align}
\end{theorem}
}

The proof of Theorem \ref{wang:mary} follows \cite{hsu2020binary} and is provided in Sections \ref{Mseqach} and \ref{Mseqcon}.

Theorems \ref{gut:mary} and \ref{wang:mary} show that the sequential test in~Eq. \eqref{sequential_universal} achieves \emph{different} error exponent under each hypothesis while Gutman's test achieves the same error exponents under each hypothesis with an additional reject option. Furthermore, Haghifam, Tan and Khisti~\cite[Theorem 6]{mahdi2021sequential} demonstrated that the achievable Bayesian error exponent of the sequential scheme is equal to that of the Gutman's test. Note that the sequential test in~Eq. \eqref{seqtest} \emph{does not} need the additional reject option used by the fixed-length Gutman's test to achieve the same Bayesian error exponent.

We wonder whether we could achieve the performance close to the optimal sequential test $\Phi_{\mathrm{seq}}^{(M)}$ with a test that does not need a reject option and has simple design mimicking Gutman's test. In the next section, we propose such a test and prove its desired performance.

\section{Main Results}
\subsection{Our Two-phase Test}
We need the following definitions to present our test. Given training sequences $\bX^N$ and a test sequence $Y^n$ and any $\alpha\in\bbR_+$, define the following function that denote the indices of the minimum scoring values
\begin{align}
i^*(\bX^N,Y^n,\alpha)&:=\mathop{\arg\min}\limits_{i\in[M]} {\rm{GJS}}\Big(\hatT_{X_i^N},\hatT_{Y^n},\alpha \Big),
\end{align}
Given $\mu\in[M]$, define the set
\begin{align}
\calI(\mu):=[M]\setminus\{\mu\}.
\end{align}

Consider any integer $n\in\bbN$ and any positive real numbers $(k,\lambda_1,\ldots,\lambda_M)\in\bbR_+^{M+1}$, our test proceeds with two phases. Specifically, the random stopping time $\tau^{(M)}$ of our test satisfies
\begin{align}\label{tau:mary}
\!\!\tau^{(M)}=\left\{\!
\begin{array}{cl}
n&\mathrm{if}~\forall i\in\calI\big(i^*(\bX^N,Y^n,\alpha)\big),~\mathrm{GJS}\Big(\hatT_{X_i^N},\hatT_{Y^n},\alpha\Big)\ge\lambda_i,\\
kn&\mathrm{otherwise}.
\end{array}
\right.
\end{align}

The decision rule $\phi_\tau$ of our test applies nearest neighbor (NN) detection based on GJS divergence using the empirical distributions of the training and testing sequence. Specifically, when $\tau=n$, our test $\phi_n$ favors hypothesis $\rmH_j$ if the divergence ${\rm{GJS}}\big(\hatT_{X_i^N},\hatT_{Y^n},\alpha \big)$ is smallest among all $\big\{{\rm{GJS}}\big(\hatT_{X_i^N},\hatT_{Y^n},\alpha \big)\big\}_{i\in[M]}$, i.e.,
\begin{align}\label{Mary:n}
\phi_n^{(M)}(\bX^N,Y^n)=\mathrm{H}_j,~ j=i^*(\bX^N,Y^n,\alpha),
\end{align}
when $\tau=kn$, we use the same test with the only exception that the empirical distribution $\hatT_{Y^{kn}}$ of all $kn$ testing sequences is used instead of $\hatT_{Y^n}$, i.e.,
\begin{align}\label{Mary:kn}
\phi_{kn}^{(M)}(\bX^N,Y^{kn})=\mathrm{H}_j,~ j=i^*(\bX^N,Y^{kn},\alpha).
\end{align}

For simplicity, we use $\Phi_\mathrm{tp}^{(M)}$ to denote the above two-phase test of $M$-ary classification. Note that our setup does not include the rejection option in the final decision.
The excess-length probability and the random stopping time for our two-phase test satisfies $\bbP_i\{\tau>n\}\leq \exp(-n\gamma)$ and $\tau\leq kn$ for some $(\gamma,k)\in\bbR^2_+$. Asymptotically, such a test ensures that the average stopping time satisfies $\limsup_{n\to\infty}\frac{\mathbb{E}[\tau]}{n}\leq 1$.

We then discuss the asymptotic performance of our two-phase test. Intuitively, for each $j\in[M]$, if the testing sequence $Y^n$ is generated from the distribution $P_j$, the empirical distribution of $\hatT_{Y^n}$ tends to $P_j$ as the length of sequence $n$ increases. Similarly, for each $j\in[M]$, $\hatT_{X_j^N}$ tend to $P_j$. Thus, under hypothesis $\rmH_j$, the scoring function $\mathrm{GJS}(\hatT_{X_j^N},\hatT_{Y^n},\alpha)$ tends to zero and all other $M-1$ scoring functions $\mathrm{GJS}(\hatT_{X_i^N},\hatT_{Y^n},\alpha),i\in\calM_j$ tend to the positive number $\mathrm{GJS}(P_i,P_j,\alpha)$. With high probability, a correct decision can be made in the first phase when $n$ is large. In an exponentially rare case where a reject is decided in the first phase, the decision maker can collect extra $(k-1)n$ samples for $k$ large and make a reliable choice in the second phase with the same logic.

\subsection{Main Results and Discussions}
For ease of notation, let $\bQ=(Q_1,Q_2,Q_3)\in\calP(\calX)^3$. Given any thresholds $\lambda^M=(\lambda_1,\ldots,\lambda_M)\in\bbR_+^M$, for each $j\in [M]$, define the following exponent function
\begin{align}\label{def:Fj}
F_j(\alpha,\lambda^M|\bP):=\min\limits_{(i,l)\in\calM_{\mathrm{dis}}}\min\limits_{\substack{\bQ\in\calP(\calX)^3:\\\mathrm{GJS}(Q_1,Q_3,\alpha)\le\lambda_i\\\mathrm{GJS}(Q_2,Q_3,\alpha)\le\lambda_l}} D(\alpha,\bQ|P_i,P_l,P_j),
\end{align}
where $D(\alpha,\bQ|P_i,P_l,P_j)=\alpha D(Q_1||P_l)+\alpha D(Q_2||P_i)+D(Q_3||P_j)$ is the linear combination of three KL divergence terms. As we shall show, $F_j(\alpha,\lambda^M|\bP)$ is critical to bound the probability that our test proceeds the second phase, i.e., $\Pr\{\tau>n\}\leq \exp(-n\gamma)$ for some targeted exponent $\gamma\in\bbR_+$. \blue{The feasible region $\calV(\bP)$ of the parameters $\lambda^M\in\bbR_+^M$ for the optimization problem in \eqref{def:Fj} satisfies
\begin{align}
\calV(\bP)=\{\lambda^M\in\bbR_+^M:~\exists~ \bQ\in\calP(\calX)^3~\mathrm{and}~\exists~(i,l)\in\calM_{\mathrm{dis}}~\mathrm{s.t.}~\mathrm{GJS}(Q_1,Q_3,\alpha)\le\lambda_i~\mathrm{and}~\mathrm{GJS}(Q_2,Q_3,\alpha)\le\lambda_l\}.
\end{align}}
For each $j\in[M]$, $F_j(\alpha,\lambda^M|\bP)$ is convex and non-increasing in $\lambda_j$. In particular, $F_j(\lambda^M|\bP)=0$ if $\lambda^M=(\lambda_1,\ldots,\lambda_M)$ satisfies that $\lambda_j>\min_{t\in\calM_j}\mathrm{GJS}(P_t,P_j,\alpha)$ for some $j\in[M]$. If $\lambda^M$ does not belong to the feasible region, i.e., $\lambda^M\notin\calV(\bP)$, the stopping time will always be $n$ (cf. \eqref{tau:mary}) and the two-phase test will reduce to a fixed-length test~\cite{gutman1989asymptotically}. Specifically, when $\lambda^M$ is a all zero vector, $F_j(\alpha,\lambda^M|\bP)$ achieves the following maximum value:
\begin{align}
\bar{\gamma}_j(\alpha|\bP)
:=\min_{(i,l)\in\calM_{\mathrm{dis}}}\min_{Q\in\calP(\calX)}D(\alpha,[Q,Q,Q]|P_i,P_l,P_j).
\end{align}

Finally, for each $j\in[M]$, define another exponent function
\begin{align}\label{def:Lj}
L_j(k,\alpha|\bP):=\min\limits_{i\in\calM_j}\min\limits_{\substack{\bQ\in\calP(\calX)^3:\\\mathrm{GJS}(Q_1,Q_3,\alpha)<\mathrm{GJS}(Q_2,Q_3,\alpha)}} U(k,\alpha,\bQ|P_i,P_j),
\end{align}
where $U(k,\alpha,\bQ|P_i,P_j)=\alpha D(Q_1||P_i)+\alpha D(Q_2||P_j)+kD(Q_3||P_j)$ is analogous to $D(\alpha,\bQ|P_i,P_l,P_j)$. As we shall show, $L_j(k,\alpha|\bP)$ characterizes the achievable type-$j$ error exponent in the second phase of our test.

With these definitions, our result states as follows.
\begin{theorem}\label{M_exponents}
For any $(k,\gamma)\in\bbR_+^2$, under any tuple of distributions $\bP\in\calP(\calX)^M$, the achievable type-$j$ error exponent of our two-phase test satisfies that for each $j\in[M]$,
\begin{align}
E_j\big(\Phi_{\mathrm{tp}}^{(M)}|\bP\big)\ge\min\{\lambda_j,L_j(k,\alpha|\bP)+\gamma\},
\end{align}
where the thresholds $\lambda^M=(\lambda_1,\ldots,\lambda_M)$ satisfy $\lambda^M\in\tilG(\gamma):=\{ \bar{\lambda}^M\in\bbR_+^M:~\min\limits_{j\in[M]}F_j(\alpha,\bar\lambda^M|\bP)\ge\gamma\}$. Furthermore, the Bayesian error exponent of our two-phase test satisfies
\begin{align}\label{M_utility}
E_\mathrm{Bayesian}\big(\Phi_{\mathrm{tp}}^{(M)}|\bP\big)\geq\max\limits_{\lambda^M\in\tilG(\gamma)}\min\limits_{j\in[M]}\min\{\lambda_j,L_j(k,\alpha|\bP)+\gamma\}.
\end{align}
\end{theorem}
The achievability proof of Theorem \ref{M_exponents} is provided in Section \ref{Maryach}. We make several remarks.

\begin{figure}[htbp]
\centering
\includegraphics[width=.5\columnwidth]{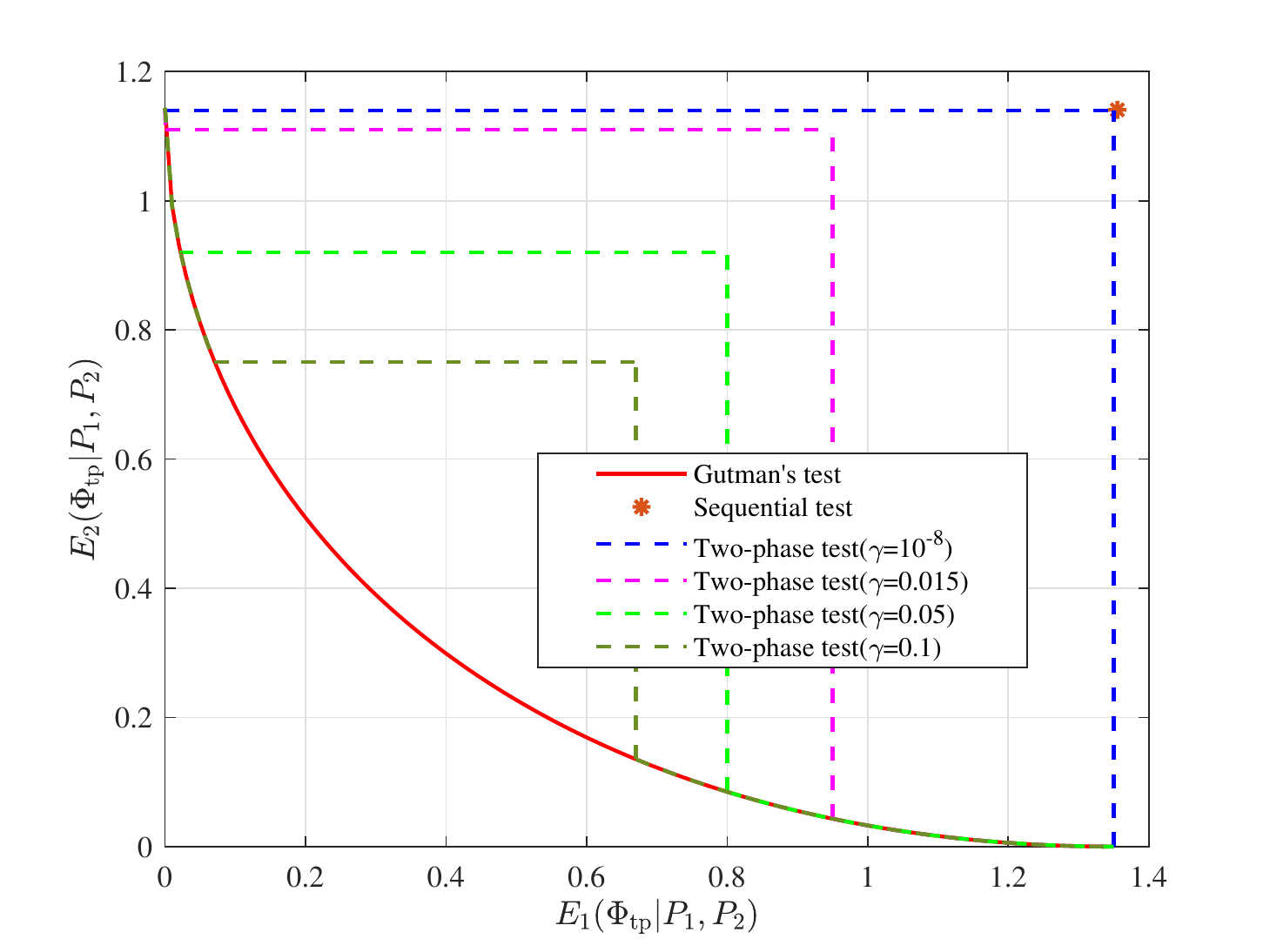}
\caption{Illustration of the type-I and type-II error exponents of $M$-ary classification under generating distributions $P_1 = [0.9,0.1]$ and $P_2 = [0.2,0.8]$ with $k=20$ and $\alpha=300$.}
\label{neyman}
\end{figure}

In the Neyman-Pearson setting, our two-phase test bridges over the fixed-length test by Gutman~\cite{gutman1989asymptotically} and the sequential test (cf. \eqref{sequential_universal}) by having error exponents close to either one with proper choices of parameters $\gamma$ and $k$. Specifically, under any tuple of distributions $\bP$, if the target value $\gamma$ satisfies $\gamma>\min\limits_{j\in[M]}\bar{\gamma}_j(\alpha|\bP)=:\underline{\gamma}(\bP,\alpha)$, the set $\tilG(\gamma)$ contains only the all zero vector, and thus the only valid thresholds $\lambda^M$ of our test are all zero. In this case, our two-phase test reduces to a fixed-length test using $n$ samples (cf. \eqref{tau:mary}). In another extreme when $\gamma\to 0$, the maximal threshold $\lambda_j$ tends to $\min_{t\in\calM_j}\mathrm{GJS}(P_t,P_j,\alpha)$. With proper choice of $k$, the achievable type-$j$ exponent of our test approach that of the optimal sequential test.
That is because, there always exists a finite $k^*\in\bbR_+$ such that $L_j(k^*,\alpha|\bP)\ge\lambda_j$ since the error exponent function $L_j(k,\alpha|\bP)$ increases linearly in $k$. Then the lower bound of the type-$j$ error exponent becomes $\min\{\lambda_j,L_j(k,\alpha|\bP)+\gamma\}=\lambda_j$. Thus the type-$j$ error exponent for the two-phase test achieves $\min_{t\in\calM_j}\mathrm{GJS}(P_t,P_j,\alpha)$ by taking $\lambda_j=\min_{t\in\calM_j}\mathrm{GJS}(P_t,P_j,\alpha)$.
For $\gamma\in(0,\underline{\gamma}(\bP,\alpha))$, our two-phase test has performance in between the fixed-length and the sequential tests. To illustrate our results, we run numerical examples to plot the achievable type-I and type-II error exponents of two-phase test with various values of $\gamma$ when $M=2$ in Fig. \ref{neyman}.

\begin{figure}[tb]
\centering
\includegraphics[width=.5\columnwidth]{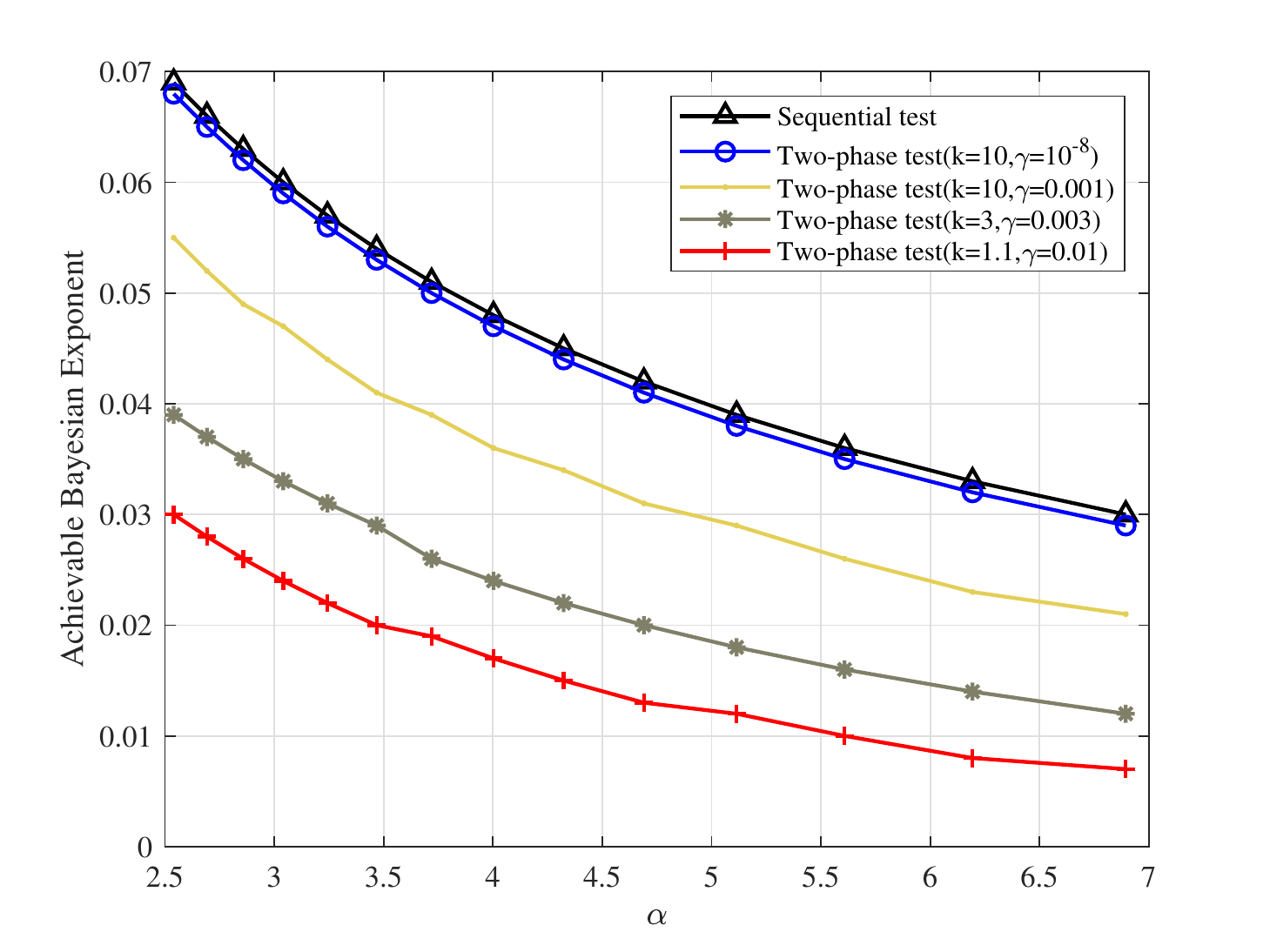}
\caption{Illustration of the Bayesian error exponent of $M$-ary classification under generating distributions $P_1 =[0.3, 0.3, 0.4]$, $P_2 =[0.4, 0.5, 0.1]$ and $P_3 =[0.1, 0.7, 0.2]$. }
\label{M_compare}
\end{figure}

In the Bayesian setting, the sequential test $\Phi_\mathrm{seq}^{(M)}$ and the fixed-length test $\Phi_{\mathrm{Gut}}^{(M)}$ achieves the same Bayesian error exponent for $M$-ary classification~\cite[Theorem 6]{mahdi2021sequential}. However, the fixed-length test  $\Phi_{\mathrm{Gut}}^{(M)}$ requires an additional reject option. In contrast, with proper choices of design parameters, our two-phase test could achieve roughly the same Bayesian error exponent without a reject option at the asymptotic complexity of the fixed-length test  $\Phi_{\mathrm{Gut}}^{(M)}$. To illustrate our result, in Fig. \ref{M_compare}, we plot the achievable Bayesian error exponents of our two-phase test and the sequential test. Consistent with \cite[Eq. (14)]{mahdi2021sequential}, the denominator in Eq. \eqref{Beyasianexponent} is set to $N$ in the numerical simulation.

Furthermore, we discuss the influence of $\alpha$, the ratio between the lengths of training sequences and the length of the test sequence on the performance of our two-phase test. For $j\in[M]$, both $F_j(\alpha,\lambda^M|\bP)$ and $L_j(k,\alpha|\bP)$ increase in $\alpha$ and thus the more the training sequences (larger $\alpha$), the better the performance. Consider the extreme case of $\alpha\to\infty$. In this case, the training sequence is unlimited so that generating distributions $\bP$ are estimated accurately and the statistical classification problem reduces to the hypothesis testing problem. The specialization to this case is available in Section \ref{sec_MHT}.

\blue{
Finally, we find it is challenging to generalize the converse analyses of Lalitha and Javidi for binary hypothesis testing to $M$-ary classification. This is mainly because when the generating distribution under each hypothesis is unknown, the optimality criterion changes dramatically. To demonstrate the optimality of our test, we derive a converse result for sequential classification under the error probability universality constraint using similar techniques as~\cite{Ihwang2022sequential}. Our two-phase test is a special sequential test satisfying the error probability universality constraint since the error probability under each hypothesis tends to $0$ when the average stopping time tends to infinity. In the above discussions, we have compared our two-phase test with Gutman's fixed-length test in Eq. \eqref{Gutman} and the sequential test under the error probability universality constraint in Eq. \eqref{seqtest}, and demonstrated that the performance of the optimal sequential test can be approached by our two-phase test with appropriate parameters.
}

\subsection{Numerical Example}
\begin{figure}[htbp]
\centering
\includegraphics[width=.5\columnwidth]{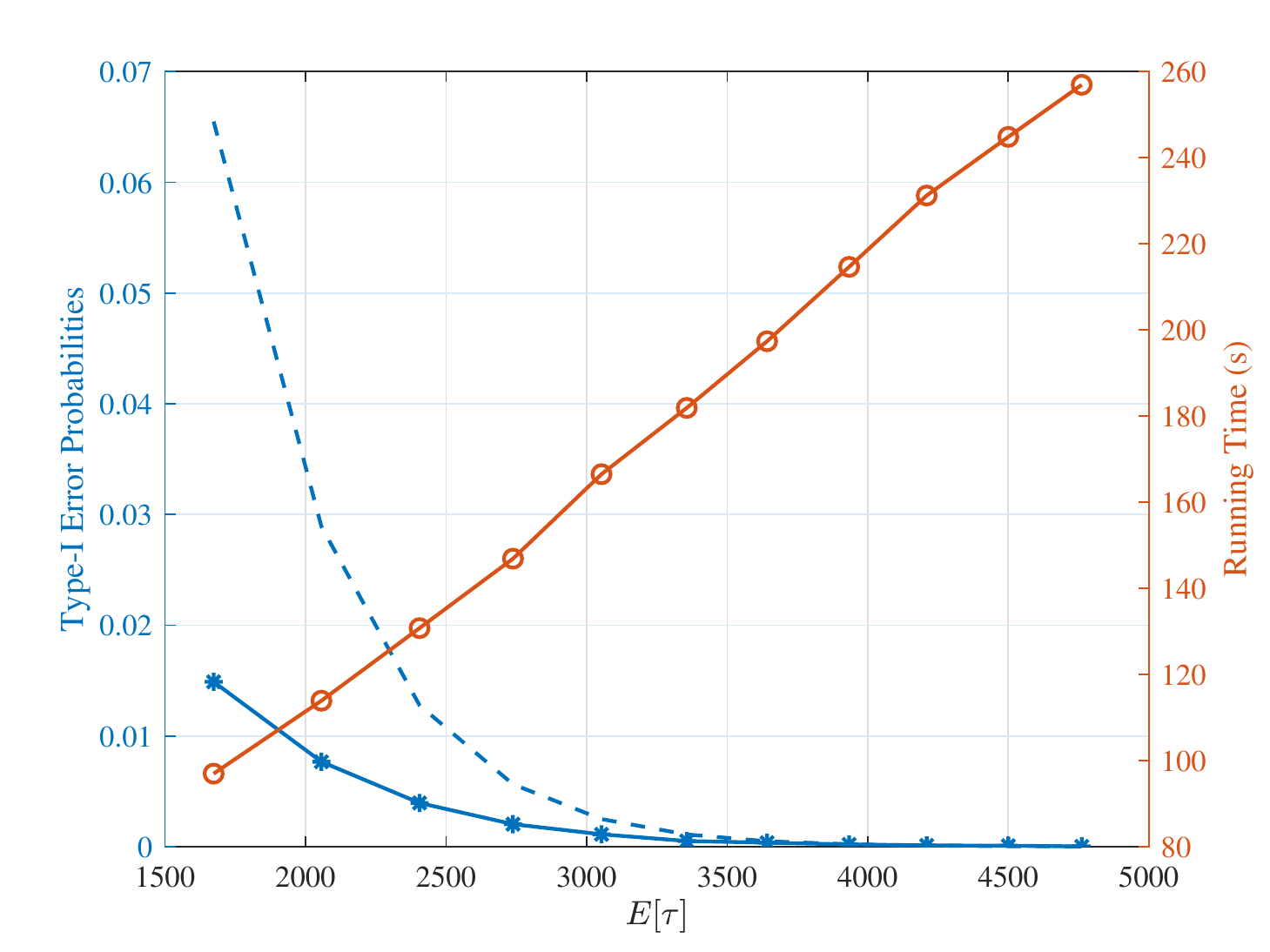}
\caption{Simulated type-I error probabilities for $M$-ary classification under generating distributions $P_1=[0.25,0.75]$, $P_2=[0.28,0.72]$ and $P_3=[0.22,0.78]$ with $k=3,\lambda_1=0.004,\lambda_2=0.0008$ and $\lambda_3=0.0009$.}
\label{simulation}
\end{figure}

\begin{figure}[htbp]
\centering
\includegraphics[width=.5\columnwidth]{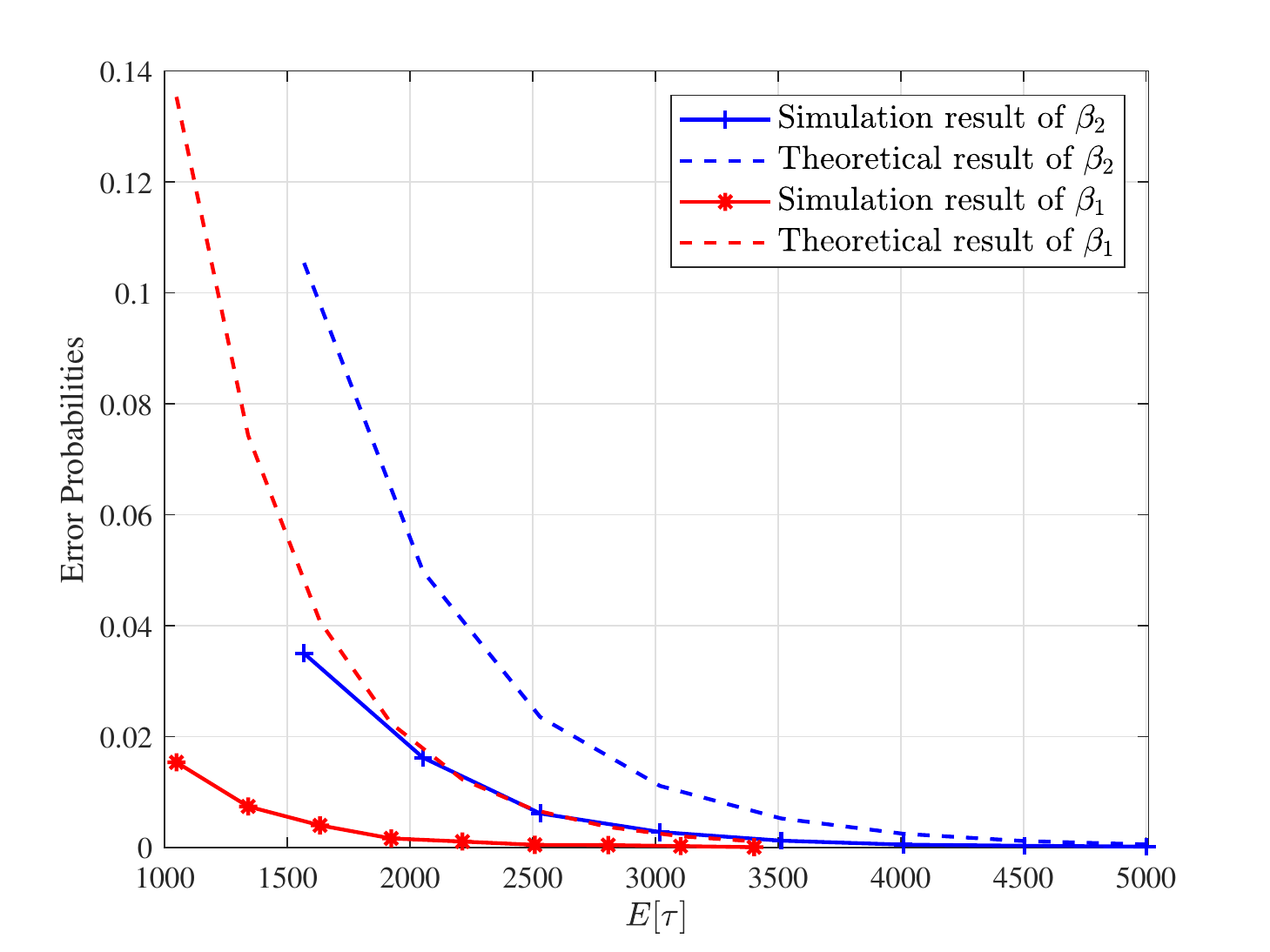}
\caption{Simulated type-I and type-II error probabilities for $P_1=[0.25,0.75]$ and $P_2=[0.2,0.8]$ with $k=3,\lambda_1=0.002,\lambda_2=0.0015$ and $\lambda=0.003$.}
\label{simulation_bin}
\end{figure}

We present a numerical example to illustrate Theorem \ref{M_exponents}. Consider the binary alphabet $\calX=\{0,1\}$ and let $k=3,\alpha=8$. In Fig. \ref{simulation}, we consider $M=3$ and plot the simulated type-I error probabilities of the two-phase test versus the exponential estimates in Theorem \ref{M_exponents} on the left y-axis and plot the running time of the two-phase test on the right y-axis. The generating distributions are $P_1=[0.25,0.75],P_2=[0.28,0.72],P_3=[0.22,0.78]$. We choose the parameters of the test as $\lambda_1=0.004$, $\lambda_2=0.0008$ and $\lambda_3=0.0009$.  For each $n\in\{1000,1500,\ldots,4000\}$, we estimate the type-I error probability of a single two-phase test using $5\times 10^{4}$ independent experiments. In Fig. \ref{simulation_bin}, we consider $M=2$ and plot the simulated type-I and type-II error probabilities in a similar manner. As observed, when $n$ is large, our theoretical characterization provides a tight upper bound. We remark that the distributions chosen here are for the calculation of error probabilities and illustration of our results. Similar results can be obtained for any tuple of distributions.

\blue{
\section{Specialization to Binary classification}
}
\subsection{Results}
We need the following definitions to present our results.
Given any $\alpha\in\bbR_+$ and any $(P_1,P_2)\in\calP(\calX)^2$, for each $i\in[2]$, let $\bari=3-i$ and $\bQ=(Q_1,Q_2,Q_3)\in\calP(\calX)^3$, define the following exponent function
\begin{align}\label{binary:F}
F_i(\alpha,\lambda_{\bari}|P_1,P_2):=\min_{\substack{(Q_1,Q_2)\in\calP(\calX)^2:\\\mathrm{GJS}(Q_{\bari},Q_i,\alpha)\le\lambda_{\bari}}} \alpha D(Q_{\bari}||P_{\bari})+D(Q_i||P_i).
\end{align}
As we shall show, $F_i(\alpha,\lambda_{\bari}|P_1,P_2)$ is critical to bound the probability that our test proceeds the second phase, i.e., $\Pr\{\tau>n\}$. Note that for each $i\in[2]$, $F_i(\alpha,\lambda_{\bari}|P_1,P_2)$ decrease in $\lambda_{\bari}$. In particular, $F_i(\alpha,\lambda_{\bari}|P_1,P_2)=0$ if $\lambda_{\bari}\geq \mathrm{GJS}(P_{\bari},P_i,\alpha)$ and achieves the following maximum when $\lambda_{\bari}=0$
\begin{align}
\min_{Q\in\calP(\calX)}\alpha D(Q\|P_{\bari})+D(Q\|P_i)=&-(1+\alpha)\log\Big[\sum\limits_{x} P_{\bari}(x)^{\frac{\alpha}{1+\alpha}}P_{i}(x)^{\frac{1}{1+\alpha}}\Big]\\*
=&D_{\frac{\alpha}{1+\alpha}}(P_{\bari}\|P_i),
\end{align}
which is the R\'enyi divergence of order $\frac{\alpha}{1+\alpha}$~\cite[Eq. (3.16)]{zhou2020second}.

Furthermore, given any $k\in\bbR_+$, let
\begin{align}\label{binary:H}
H_i(k,\alpha|P_1,P_2):=\min_{\substack{\bQ\in\calP(\calX)^3:\\ \mathrm{GJS}(Q_{\bari},Q_3,\alpha)<\mathrm{GJS}(Q_i,Q_3,\alpha)}}\alpha D(Q_1||P_1)+\alpha D(Q_2||P_2)+kD(Q_3||P_i).
\end{align}
As we shall show, $H_i(k,\alpha|P_1,P_2)$ characterizes the error exponent in the second phase of our test where we use a NN test with $kn$ samples.

With these definitions, our main result states as follows.
\begin{theorem}\label{two_exponents}
For any $n\in\bbN,~k\in\bbR_+$ and $\gamma\in\bbR_+$, under any pair of distributions $(P_1,P_2)$, the achievable type-I and type-II error exponents of our two-phase test satisfy
\begin{align}
&E_1(\Phi_{\mathrm{tp}}|P_1,P_2)\ge\min\big\{\lambda_1,H_1(k,\alpha|P_1,P_2)+\gamma\big\},\\
&E_2(\Phi_{\mathrm{tp}}|P_1,P_2)\ge\min\big\{\lambda_2,H_2(k,\alpha|P_1,P_2)+\gamma\big\},
\end{align}
where the thresholds $(\lambda_1,\lambda_2)$ satisfy $(\lambda_1,\lambda_2)\in G(\gamma):=\{(\bar{\lambda}_1,\bar{\lambda}_2)\in\bbR_+^2:~\min\limits_{i\in[2]}F_i(\alpha,\bar{\lambda}_1,\bar{\lambda}_2|P_1,P_2)\ge\gamma\}$ and $\lambda_2\neq0$. When $\lambda_2=0$, the two-phase test reduces to Gutman's test and thus the achievable error exponents reduce to Gutman's~\cite[Theorem 3]{gutman1989asymptotically}.
Furthermore, the achievable Bayesian error exponent of our two-phase test satisfies under any pair of distributions $(P_1,P_2)$,
\begin{align}\label{E}
E_{\mathrm{Bayesian}}(\Phi_{\mathrm{tp}}|P_1,P_2)\geq \max\limits_{(\lambda_1,\lambda_2)\in G(\gamma)}\min\limits_{i\in[2]}\min\big\{\lambda_i,H_i(k,\alpha|P_1,P_2)+\gamma\big\}.
\end{align}
\end{theorem}

We remark that the proof of Theorem \ref{two_exponents} is similar to binary classification with a reject option~\cite{zhou2020second}. The probability that our test enters a second phase equals the probability that a reject option is decided in binary classification, which implies that further investigation is needed. The same probability was studied in the second-order asymptotic regime in \cite{zhou2020second} while we study the large deviations regime. The derivations of the error exponents in both phases of our test follow Gutman~\cite{gutman1989asymptotically}.

We also make several other remarks. Firstly, Theorem \ref{two_exponents} generalizes the results of Lalitha and Javidi~\cite{lalithaalmost} for binary hypothesis testing to binary classification where generating distributions are unknown under each hypothesis. In the same spirit of \cite{lalithaalmost}, our two-phase test bridges over the fixed-length test by Gutman~\cite{gutman1989asymptotically} and the sequential test by Hsu \emph{et al.}~\cite{Ihwang2022sequential} by having error exponents close to either one with proper choices of parameters $\gamma$. Specifically, under any pair of distributions $(P_1,P_2)$, if the target value $\gamma$ satisfies $\gamma>\min\{\underline{\gamma}_1(P_1,P_2,\alpha),\underline{\gamma}_2(P_1,P_2,\alpha)\}=:\underline{\gamma}(P_1,P_2,\alpha)$, we have $\lambda_2=0$ and our test reduces to Gutman's test. This is because the maximum exponent of the excess-length probability $\Pr\{\tau>n\}$ that our two-phase test can ensure is $D_{\frac{\alpha}{1+\alpha}}(P_{\bari}\|P_i)$ under hypothesis $\rmH_i$. In another extreme when $\gamma\to 0$, $\lambda_i$ tends to $\mathrm{GJS}(P_i,P_{\bari},\alpha)$. With proper choice of $k$, the achievable exponents of our test approach the exponents of the sequential test in \cite{Ihwang2022sequential}. For $\gamma\in(0,\underline{\gamma}(P_1,P_2,\alpha))$, our two-phase test has performance in between the fixed-length and the sequential tests.

\subsection{Comparison to Another Test}
\blue{
We next recall the threshold-based two-phase test in our previous conference paper~\cite[Eq. (14) and Eq. (15)]{bai2022achievable}, which also consists of a random stopping time and decision rules in two phases. The random stopping time $\tau$ is
\begin{align}\label{tau:binary}
\tau=\left\{
\begin{array}{cl}
n&\mathrm{if}~\exists~i\in[2],~\mathrm{GJS}\big(\hatT_{X_i^N},\hatT_{Y^n},\alpha\big)>\lambda_i,\\
kn&\mathrm{otherwise}.
\end{array}
\right.
\end{align}
When $\tau=n$ and $\tau=kn$, the decision rules are two fixed-length Gutman's tests with thresholds $(\lambda_1,\lambda)\in\bbR_+^2$ respectively as follows:
\begin{align}
\phi_n(X_1^N,X_2^N,Y^n)=\left\{
\begin{array}{l}
\mathrm{H_1}\;\;\mathrm{if}~\mathrm{GJS}\big(\hatT_{X_1^N},\hatT_{Y^n},\alpha\big)\leq\lambda_1,\\
\mathrm{H_2}\;\;\mathrm{otherwise}.
\end{array}\right.
\end{align}
\begin{align}
\phi_{kn}(X_1^N\!,X_2^N,Y^{kn})=\left\{\begin{array}{l}
\mathrm{H_1}\;\;\mathrm{if}\;\mathrm{GJS}\big(\hatT_{X_1^N},\hatT_{Y^{kn}},\frac{\alpha}{k}\big)\le\lambda,\\
\mathrm{H_2}\;\;\mathrm{otherwise}.
\end{array}\right.
\end{align}
}

\blue{
In~\cite[Theorem 3]{bai2022achievable}, we derive the achievable error exponents of the above threshold-based two phase test. To present the result, we need the following definitions. Given any $\alpha\in\bbR_+$ and any $(P_1,P_2)\in\calP(\calX)^2$, for each $i\in[2]$, let $\bari=3-i$ and define
\begin{align}
\nn F_i(\alpha,\lambda_{\bari}|P_1,P_2):=\min_{\substack{(Q_1,Q_2)\in\calP(\calX)^2:\\\mathrm{GJS}(Q_{\bari},Q_i,\alpha)\le\lambda_{\bari}}} \alpha D(Q_{\bari}||P_{\bari})+D(Q_i||P_i).
\end{align}
Furthermore, given any $k\in\bbR_+$, let
\begin{align}
\nn H(k,\alpha,\lambda|P_1,P_2):=\min_{\substack{(Q_1,Q_2)\in\calP(\calX)^2\\ \mathrm{GJS}(Q_1,Q_2,\alpha)\le\lambda}}\alpha D(Q_1||P_1)+kD(Q_2||P_2).
\end{align}
\begin{theorem}
For any $n\in\bbN,~k\in\bbR_+$ and $\gamma\in\bbR_+$ and any pair of distributions $(P_1,P_2)$, the achievable type-I and type-II error exponents of the threshold-based two-phase test satisfy
\begin{align}
&E_1(\Phi_{\mathrm{tp}}|P_1,P_2)\ge\min\big\{\lambda_1,k\lambda+\gamma\big\},\\
&E_2(\Phi_{\mathrm{tp}}|P_1,P_2)\ge\min\big\{\lambda_2,H(k,\alpha,\lambda|P_1,P_2)+\gamma\big\},
\end{align}
where $(\lambda_1,\lambda_2)\in\bbR_+$ satisfy $\min\limits_{i\in[2]}F_i(\alpha,\lambda_{\bari}|P_1,P_2)\ge\gamma$ and $\lambda_2\neq0$. When $\lambda_2=0$, the two-phase test reduces to Gutman's test and thus the error exponents.
\end{theorem}
}

\begin{figure}[tb]
\centering
\includegraphics[width=.5\columnwidth]{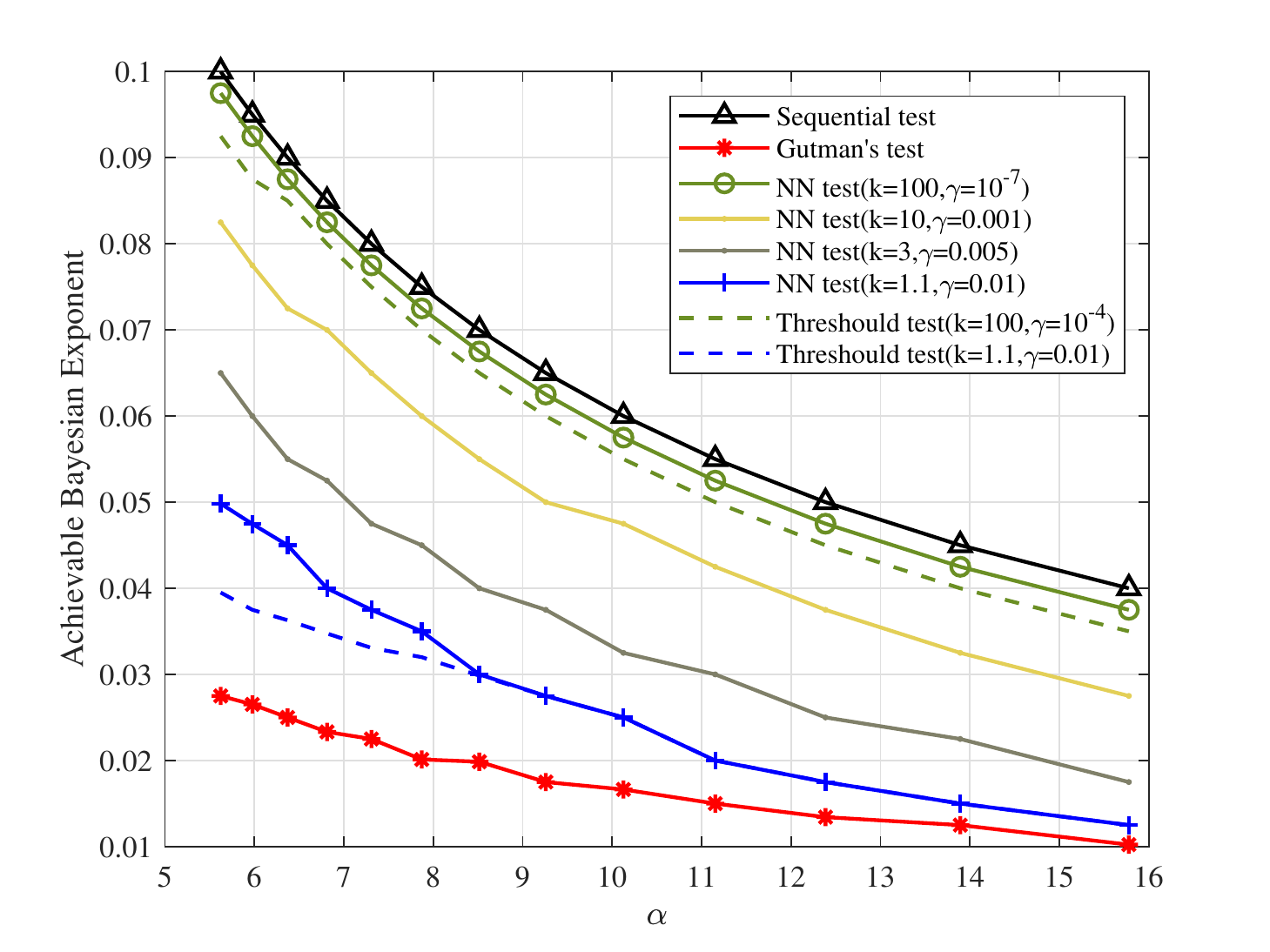}
\caption{Illustration of the Bayesian error exponent of binary classification under generating distributions $P_1= [0.1,0.3,0.6]$ and $P_2= [0.45,0.45,0.1]$.}
\label{bayesian_exponent}
\end{figure}

\blue{
For the binary case, we compare the performance of our NN-based two-phase test with the above threshold-based two-phase test, the fixed-length Gutman's test in~\cite{gutman1989asymptotically} and the sequential test in~\cite[Theorem 2]{Ihwang2022sequential} via numerical results. In Fig. \ref{bayesian_exponent}, we plot the Bayesian error exponent of these tests with different parameters, which demonstrate that our NN-based test is superior to the threshold-based test and also bridges over Gutman's test and the sequential test. The extreme case of $k=1$ and appropriately chosen $\gamma^*$ corresponds to a  fixed-length test and another extreme case of $\gamma\to 0$ and $k$ sufficient large corresponds to a sequential test.
}

\blue{
We next discuss the influence of $\alpha$, the ratio between the lengths of training sequences and the length of the testing sequence on the performance of the threshold-based two-phase test. Both $F_i(\alpha,\lambda_{\bari}|P_1,P_2),i\in[2]$ and $H_i(k,\alpha,\lambda|P_1,P_2)$ increase in $\alpha$ and thus the more the training sequences (larger $\alpha$), the better the performance. We first consider the extreme case of $\alpha\to\infty$, i.e., when the training sequence is unlimited so that generating distributions $(P_1,P_2)$ are estimated accurately, for each $i\in[2]$, we have
\begin{align}
F_i(\infty,\lambda_{\bari}|P_1,P_2)=\min\limits_{\substack{Q_i\in\calP(\calX):\\D(Q_i||P_{\bari})\le\lambda_{\bari}}}D(Q_i||P_i)=D(Q_i^*||P_i),
\end{align}
where $Q_i^*$ satisfies
\begin{align}
Q_i^*= \frac{P_i(x)^{\frac{1}{1+r^*_i}}P_{\bari}(x)^{\frac{r^*_i}{1+r^*_i}}}{\sum\limits_{a\in\calX}P_i(x)^{\frac{1}{1+r^*_i}}P_{\bari}(x)^{\frac{r^*_i}{1+r^*_i}}},
\end{align}
and $(r^*_1,r^*_2)$ are Lagrange multipliers such that $D(Q^*_1||P_2)=\lambda_2$ and $D(Q^*_2||P_1)=\lambda_1$ respectively. Furthermore, we have
\begin{align}
\nn H(k,\infty,\lambda|P_1,P_2)=\min\limits_{\substack{Q\in\calP(\calX):\\D(Q||P_1)\le\lambda}}kD(Q||P_2)=kD(Q^\prime||P_2).
\end{align}
where $Q^\prime$ satisfies
\begin{align}
Q^\prime=\frac{P_1(x)^{\frac{r^*}{k+r^*}}P_2(x)^{\frac{k}{k+r^*}}}{\sum\limits_{a\in\calX}P_1(a)^{\frac{r^*}{k+r^*}}P_2(a)^{\frac{k}{k+r^*}}},
\end{align}
and $r^*$ is the Lagrange multiplier such that $D(Q^\prime||P_1)=\lambda$.
Then we have the type-I and type-II error exponents satisfy
\begin{align}
&E_1(\Phi_{\mathrm{tp}}|P_1,P_2)\!=\!\min\big\{D(Q_2^*||P_1),kD(Q^\prime||P_1)+\gamma\big\},\\
&E_2(\Phi_{\mathrm{tp}}|P_1,P_2)\!=\!\min\big\{D(Q_1^*||P_2),kD(Q^\prime||P_2)+\gamma\big\},
\end{align}
where $(Q_1^*,Q_2^*)$ satisfy $\min\limits_{i\in[2]}D(Q_i^*||P_i)>\gamma$. In the other extreme case of $\alpha\to 0$, we find that $H(k,\alpha,\lambda|P_1,P_2)=0$ and $F_i(\alpha,\!\lambda_{\bari}|P_1,P_2)=0$ for each $i\in[2]$.
}

\blue{
\section{Specialization to hypothesis testing}\label{sec_MHT}
}
In this section, we specialize our results to multiple hypothesis testing by letting $\alpha\to\infty$ so that the generating distribution under each hypothesis is exactly estimated. In multiple hypothesis testing, we are given an observed sequence $Y^\tau=(Y_1,\ldots,Y_\tau)\in\calX^\tau$ where $\tau$ is a random stopping time with respect to the Filtration $\sigma\{Y_1,\ldots Y_n\}$. The observed sequence is generated i.i.d. from one of $M$ distributions $\bP:=(P_1,\ldots,P_M)\in\calP(\calX)^M$ and thus there exists $M$ hypotheses $\{\mathrm{H}_j\}_{j\in[M]}$ which correspond to $M$ possible underlying known distributions. The task for multiple hypothesis testing problem is to design a test $\Phi=(\tau,\phi_\tau)$, which consists of a random stopping time $\tau$ and a mapping $\phi_\tau$: $\calX^\tau\rightarrow\{\rmH_1,\ldots,\rmH_M\}$, to decide among the following $M$ hypotheses:
\begin{itemize}
\item $\rmH_j$: the sequence $Y^\tau$ is generated i.i.d. from the distribution $P_j$,~$j\in[M]$.
\end{itemize}

\subsection{Two-phase Test}\label{MHTtest}
In this subsection, we present our two-phase test that uses the empirical distributions of the testing sequences and proceeds in two phases.
In the first phase, the test takes $n$ samples $Y^n$ to perform a fixed-length test with a reject option,  where reject means that more samples are needed to make a reliable decision. Once a reject decision is made in the first phase, our test proceeds in the second phase, where $(k-1)n$ additional test samples $Y_{n+1}^{kn}$ are collected and a fixed-length test \emph{without} a rejection is used to make a final decision.

Given any integers $(M,n)\in\bbN^2$, any tuple of distributions $\bP\in\calP(\calX)^M$ and a test sequence $Y^n$, the stopping time for $M$-ary hypothesis testing follows by specializing the stopping time for $M$-ary classification in Eq. \eqref{tau:mary} to $\alpha\to\infty$, i.e.,
\begin{align}
\label{tau:mht}
\tau=\left\{
\begin{array}{cl}
n&\mathrm{if}~~\forall~i\in\calI\big(t^*(Y^n,\bP)\big),~D(\hatT_{Y^n}\|P_i)>\lambda_i,\\
kn&\mathrm{otherwise},
\end{array}
\right.
\end{align}
where $t^*(Y^n,\bP)=i^*(\bX^N,Y^n,\infty)=\mathop{\arg\min}_{i\in[M]}D\big(\hatT_{Y^n}\|P_i\big)$.
When $\tau=n$ and $\tau=kn$, the decision rule $\phi_\tau$ for $M$-ary hypothesis testing also follows from the decision rule for $M$-ary classification in Eq. \eqref{Mary:n} and Eq. \eqref{Mary:kn} respectively, when specialized to $\alpha\to\infty$, i.e.,
\begin{align}
\label{argmin1}\phi_n(Y^n)&=\mathrm{H}_j,~j=t^*(Y^n,\bP),\\
\label{argmin2}\phi_{kn}(Y^{kn})&=\mathrm{H}_j,~j=t^*(Y^{kn},\bP).
\end{align}

For simplicity, we use $\Phi_{\rm{ht}}^{(M)}$ to denote our two-phase test for $M$-ary hypothesis testing.

\subsection{Main Results and Discussions}

Given any tuple of distributions $\bP=(P_1,\ldots,P_M)\in\calP(\calX)^M$ and any thresholds $\lambda^M=(\lambda_1,\ldots,\lambda_M)\in\bbR_+^M$, define the exponent function
\begin{align}
\Gamma_j\big(\lambda^M|\bP\big)&:=\min\limits_{(i,l)\in\calM_{\mathrm{dis}}}\min\limits_{\substack{Q\in\calP(\calX):\\D(Q\|P_i)\le\lambda_i,D(Q\|P_l)\le\lambda_l}}D(Q\|P_j), \label{Gammaj}
\end{align}
As we shall show, $\Gamma_j(\lambda^M|\bP)$ is critical to bound the exponential decaying probability that the random stopping time exceeds $n$, i.e., $\Pr\{\tau>n\}\leq \exp(-n\gamma)$ for some targeted exponent $\gamma\in\bbR_+$. \blue{The feasible region $\calL(\bP)$ of the parameters $\lambda^M\in\bbR_+^M$ for the optimization problem in \eqref{Gammaj} satisfies
\begin{align}
\calL(\bP)=\{&\lambda^M\in\bbR_+^M:\exists~ Q\in\calP(\calX)~\mathrm{and}~\exists~(i,l)\in\calM_{\mathrm{dis}}~\mathrm{s.t.}~D(Q\|P_i)\le\lambda_i~\mathrm{and}~D(Q\|P_l)\le\lambda_l\}.
\end{align}}
For each $j\in[M]$, note that $\Gamma_j(\lambda^M|\bP)$ is non-increasing in $\lambda_j$. In particular, $\Gamma_j\big(\lambda^M|\bP\big)=0$ if $\lambda^M=(\lambda_1,\ldots,\lambda_M)$ satisfies that $\lambda_j\ge\min\limits_{(i,l)\in\calM_{\mathrm{dis}}}\max\{D(P_j\|P_i),D(P_j\|P_l)\}$ for some $j\in[M]$. If $\lambda^M$ does not belong to the feasible region, i.e., $\lambda^M\notin\calL(\bP)$, the stopping time will always be $n$ (cf. \eqref{tau:mht}) and the two-phase hypothesis testing will reduce to a fixed-length hypothesis testing. Specifically, when $\lambda^M$ is a all zero vector, the maximal finite value of the right hand side of \eqref{Gammaj} over $\lambda^M\in\bbR_+^M$ is reached and denoted as $\bar{\gamma}_j(\bP)$.

Furthermore, given any $k\in\bbR^+$, define the exponent function
\begin{align}
\Omega_j\big(\bP\big):=\min\limits_{i\in\calM_j}\min\limits_{\substack{Q\in\calP(\calX):\\D(Q\|P_i)<D(Q\|P_j)}}D(Q\|P_j).
\end{align}
As we shall show, $\Omega_j(\bP)$ characterizes the $j$-th error exponent in the second phase of our test with $kn$ samples.

With these definitions, our result states as follows.
\begin{theorem}\label{MHT}
For any $(k,\gamma)\in\bbR_+^2$ and any tuple of distributions $\bP\in\calP(\calX)^M$, the achievable type-$j$ error exponents of our two-phase test satisfies that for each $j\in[M]$,
\begin{align}
E_j\big(\Phi_{\mathrm{ht}}^{(M)}|\bP\big)\ge\min\big\{\lambda_j,k\Omega_j(\bP)+\gamma\},
\end{align}
where the thresholds $\lambda^M=(\lambda_1,\ldots,\lambda_M)$ satisfy $\lambda^M\in\hatG(\gamma):=\{\bar{\lambda}^M\in\bbR_+^M:~\min_{j\in[M]}\Gamma_j\big(\bar{\lambda}^M|\bP\big)\ge\gamma\}$. Furthermore, the Bayesian error exponent of our two-phase test satisfies
\begin{align}
E_\mathrm{Bayesian}\big(\Phi_{\mathrm{ht}}^{(M)}|\bP\big)\geq\max_{\lambda^M\in\hatG(\gamma)}\min_{j\in[M]}\min\big\{\lambda_j,k\Omega_j(\bP)+\gamma\}.
\end{align}
\end{theorem}
We make several remarks.

Firstly, our results generalize the results of Lalitha and Javidi~\cite{lalithaalmost} for binary hypothesis testing to multiple hypothesis testing with more than two decision outcomes. In the same spirit of \cite{lalithaalmost}, our two-phase test bridges over the fixed-length test by Tuncel~\cite{tuncel2005error} and the sequential test by Baum and Veeravalli~\cite{baum1994sequential} since the achievable exponents of our test approach exponents of either case with proper choices of parameters $\gamma$ and $k$. Specifically, under any tuple of distributions $\bP$, if $\gamma$ is large enough so that $\gamma>\min_{j\in[M]}\bar{\gamma}_j(\bP)=:\bar{\gamma}(\bP)$, the set $\hatG(\gamma)$ contains only the all zero vector, and thus the only valid thresholds $\lambda^M$ of our test are all zero. In this case, our two-phase test reduces to a fixed-length test using $n$ samples (cf. \eqref{tau:mht}). On the other hand, if $\gamma\to 0$, for each $j\in[M]$, the maximum threshold $\lambda_j$ approaches $\min\limits_{(i,l)\in\calM_{\mathrm{dis}}}\max\{D(P_j\|P_i),D(P_j\|P_l)\}$. With proper choice of $k$, the achievable exponents of our test approaches the achievable exponent of the optimal sequential test~\cite{baum1994sequential}. Finally, if $\gamma\in\big(0,\bar{\gamma}\big(\bP\big)\big)$, our two-phase test has performance in between the fixed-length test and the sequential test. In the same logic, the achievable Bayesian exponent of our test also bridges that of a fixed-length and a sequential test by tuning parameters $\gamma$ and $k$.

\begin{figure}[tb]
\centering
\includegraphics[width=.5\columnwidth]{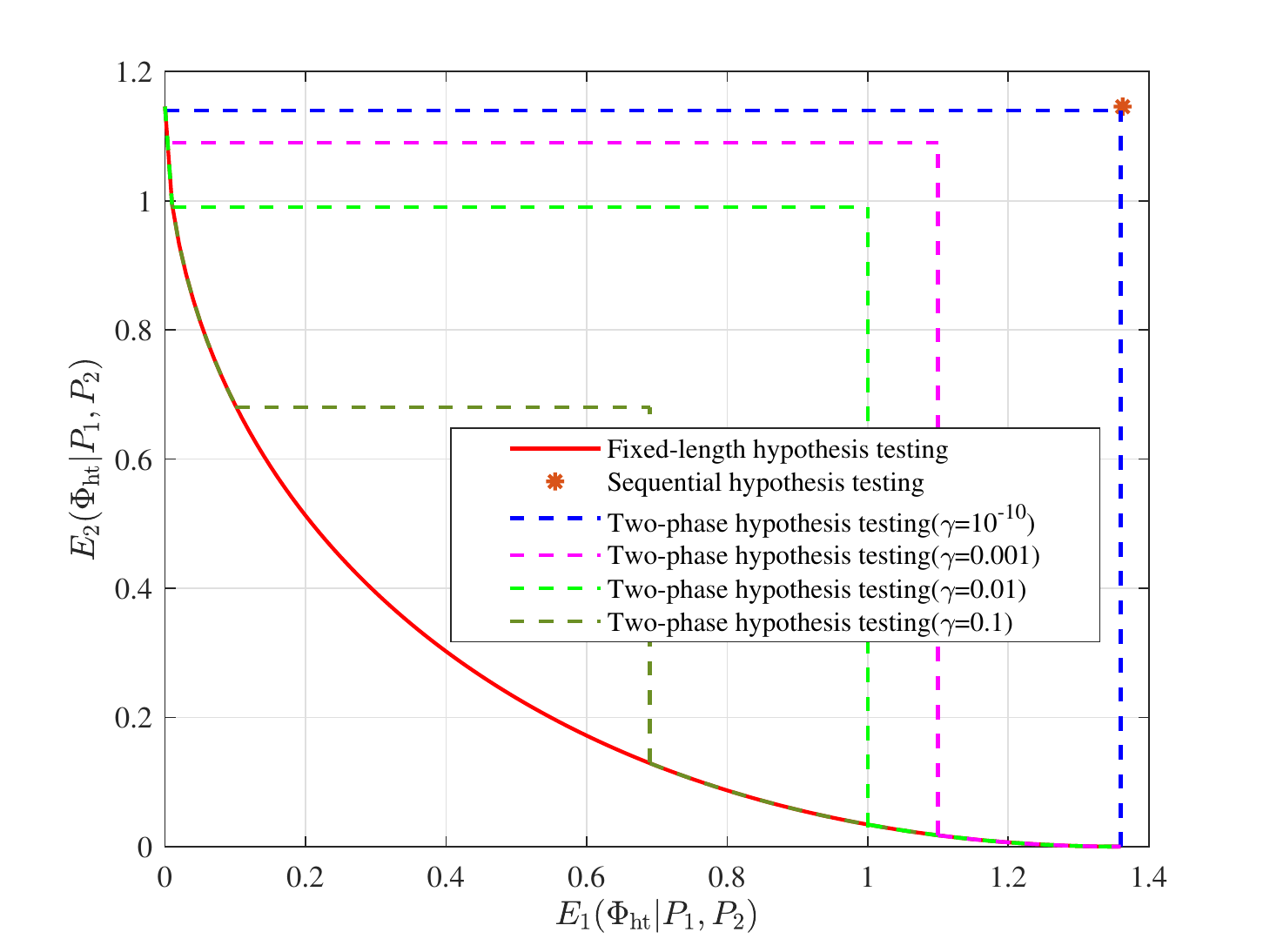}
\caption{Illustration of the type-I and type-II error exponents of multiple hypothesis testing under generating distributions $P_1 = [0.9,0.1]$ and $P_2 = [0.2,0.8]$ when $M=2$ and $k=10$.}
\label{MHT_compare}
\end{figure}

To illustrate our results, we run numerical examples to plot the achievable exponents of our two-phase test and compare our results with optimal achievable exponents of the fixed-length test~\cite{tuncel2005error} and the sequential test~\cite{baum1994sequential}. Specifically, in Fig. \ref{MHT_compare}, we plot the achievable type-I and type-II error exponents of two-phase test with various values of $\gamma$ and $k$ when $M=2$. By tuning the parameters, when $\gamma$ is large, the performance of our test approaches the fixed-length test and when $\gamma$ is small, the performance of our test approaches the sequential test as desired. Furthermore, in Table. \ref{tab:Bayesian}, we numerically compare the achievable Bayesian error exponent of our test for $M=3$ with fixed-length and sequential tests under generating distributions $P_1 = [0.3,0.3,0.4]$, $P_2 = [0.4,0.5,0.1]$ and $P_3=[0.1,0.7,0.2]$. The numerical results verify that the performance of test bridges over that of fixed-length and sequential tests by tuning the parameters of $(k,\gamma)$.

\begin{table}
\centering
\caption{Bayesian error exponents of multiple hypothesis testing.}
\label{tab:Bayesian}
\begin{tabular}{cc}
\toprule
Parameters& Bayesian error exponent\\
\midrule
Sequential hypothesis testing& 0.2319\\
Two-phase hypothesis testing\\($\gamma=10^{-8},k=10$)& 0.2310\\
Two-phase hypothesis testing\\($\gamma=0.001,k=2.2$)& 0.2013\\
Two-phase hypothesis testing\\($\gamma=0.005,k=1.8$)& 0.1647\\
Two-phase hypothesis testing\\($\gamma=0.01,k=1.1$)& 0.1007\\
Fixed-length hypothesis testing& 0.0936\\
\bottomrule
\end{tabular}
\end{table}

Finally, we remark that our two-phase test is based on the Hoeffding's test~\cite{hoeffding1965asymptotically} and is different from the test by Lalitha and Javidi~\cite{lalithaalmost} that is based on the likelihood ratio test. Specifically, the test in~\cite{lalithaalmost} compares the log likelihood ratios with different thresholds when $M=2$ while our test uses the KL divergence based nearest neighbor detection rule with thresholds. We find that our test is more amenable to be generalized to statistical classification where the generating distributions are unknown.

\section{Proofs of the main results}

\subsection{Achievability Proof of Theorem \ref{wang:mary}}\label{Mseqach}
\blue{
In the following, we define $N_k:=\lceil\alpha k\rceil$ and $k\in\bbN,~k\ge1$ to denote the sample size of each training sequence and the testing sequence at time $k$, respectively.
\subsubsection{Error Probability Universality}
We first show that the sequential test in Eq. \eqref{seqtest} satisfies the universality constraint on error probability with $\beta$. Specifically,
\begin{align}
&\bbP_j\{\phi_{\tau_\mathrm{seq}}^{(M)}(\bX^{N_{\tau_{\mathrm{seq}}}},Y^{\tau_{\mathrm{seq}}})\neq\mathrm{H}_j\}\\*
&\le\sum_{k=1}^{\infty}\bbP_j\{\phi_{k}^{(M)}(\bX^{N_k},Y^k)\neq\mathrm{H}_j\}\\
&\le\sum_{k=1}^{\infty}\bbP_j\big\{\mathrm{GJS}\big(\hatT_{X_j^{N_k}},\hatT_{Y^k},\alpha\big)>g(\beta,k)\big\}\\
&\le\sum_{k=1}^{\infty}\sum_{\substack{(Q_1,Q_2)\in\calP^{N_k}(\calX)\times\calP^k(\calX):\\\mathrm{GJS}(Q_1,Q_2,\alpha)>
g(\beta,k)}}\exp\{-N_kD(Q_1||P_j)-kD(Q_2||P_j)\}\label{expupper}\\
&\le\sum_{k=1}^{\infty}(k+1)^{|\calX|}(N_k+1)^{|\calX|}\exp\{-kg(\beta,k)\}\label{knumber}\\
&=\sum_{k=1}^{\infty}(k+1)^{|\calX|}(N_k+1)^{|\calX|}\exp\big\{\log\big(\beta(|\calX|-1)\big)-2|\calX|\log(k+1)-|\calX|\log(N_k+1)\big\}\label{gk}\\
&\le\beta(|\calX|-1)\sum_{k=1}^{\infty}(k+1)^{-|\calX|}\\
&\le\beta(|\calX|-1)\int_{0}^{\infty}(u+1)^{-|\calX|}\mathrm{d}u\label{continuous}\\
&=\beta(|\calX|-1)\frac{1}{-|\calX|+1}(u+1)^{-|\calX|+1}\Big|_{u=0}^{u=\infty}\\
&=\beta,
\end{align}
where \eqref{expupper} follows from the upper bound of the probability of a type class, \eqref{knumber} follows from the fact that the number of the set of types of length $n$ satisfies $|\calP^n(\calX)|\le(n+1)^{|\calX|}$ and the definition of GJS divergence: $\mathrm{GJS}(P,Q,\alpha)=\min_{V\in\calP(\calX)}\alpha D(P||V)+D(Q||V)$, \eqref{gk} follows from the definition of $g(\beta,k)$ in Eq. \eqref{gbeta}, \eqref{continuous} follows from the similar manner as~\cite[Appendix C: 1)]{Ihwang2022sequential}.
}

\blue{
\subsubsection{Achievable Error Exponents}
In the following, we drive the achievable error exponents for the sequential test in Eq. \eqref{seqtest} when $\beta\to 0$.
}

\blue{
Given $j\in[M]$, define the stopping time
\begin{align}\label{def:tauj}
\tau_j:=\inf\big\{k\in\bbN:\forall i\in\calM_j,\mathrm{GJS}\big(\hatT_{X_i^{N_k}},\hatT_{Y^k},\alpha\big)>g(\beta,k)\big\}.
\end{align}
We first prove $-\frac{\tau_j}{\log\beta}$ converge to $\min\limits_{i\in\calM_j}\frac{1}{\mathrm{GJS}(P_i,P_j,\alpha)}$ in probability.}
\begin{proof}
\blue{Under hypothesis $\mathrm{H}_j$, by the definition of the stopping time $\tau_j$, we have
\begin{align}
\label{ge}\min\limits_{i\in\calM_j}\mathrm{GJS}\Big(\hatT_{X_i^{N_{\tau_j}}},\hatT_{Y^{\tau_j}},\alpha\Big)>g(\beta,\tau_j).
\end{align}
}

\blue{
In the following, we show that $\tau_j\to\infty$ as $\beta\to 0$. Given $k\in\bbN$, we have
\begin{align}
\nn&\bbP_j\{\tau_j\le k\}\\*
&=\bbP_j\Big\{\forall i\in\calM_j,\mathrm{GJS}\Big(\hatT_{X_i^{N_{\tau_j}}},\hatT_{Y^{\tau_j}},\alpha\Big)>g(\beta,k),\tau_j\le k\Big\}\label{deftau}\\
&\le\max_{i\in\calM_j}\bbP_j\Big\{\mathrm{GJS}\Big(\hatT_{X_i^{N_{\tau_j}}},\hatT_{Y^{\tau_j}},\alpha\Big)>g(\beta,k),\tau_j\le k\Big\}\\
&\le\max_{i\in\calM_j}\bbP_j\Big\{k\min\limits_{V}\Big\{\alpha D\Big(\hatT_{X_i^{N_{\tau_j}}}||V\Big)+D(\hatT_{Y^{\tau_j}}||V)\Big\}>-\log\big(\beta(|\calX|-1)\big)\Big\}\label{GJS2}\\
&\le\max_{i\in\calM_j}\bbP_j\Big\{k\max\{\alpha,1\}\min\limits_{V}\Big\{ D\Big(\hatT_{X_i^{N_{\tau_j}}}||V\Big)+D(\hatT_{Y^{\tau_j}}||V)\Big\}>-\log\big(\beta(|\calX|-1)\big)\Big\}\\
&\le \bbP_j\big\{2k\max\{\alpha,1\}>-\log\big(\beta(|\calX|-1)\big)\big\},\label{Dle1}
\end{align}
where \eqref{deftau} follows from Eq. \eqref{ge}, \eqref{GJS2} follows from the definition of GJS divergence:  $\mathrm{GJS}(P,Q,\alpha)=\min_{V\in\calP(\calX)}\alpha D(P||V)+D(Q||V)$ and \eqref{Dle1} follows from the fact that the upper bound of Jesen-Shannon divergence is $1$~\cite{lin1991divergence}. Thus we can obtain that for each $j\in[M]$,
\begin{align}
\bbP_j\{\tau_j\le k\}=0,\quad \forall k<-\frac{\log\big(\beta(|\calX|-1)\big)}{2\max\{\alpha,1\}}.\label{tautoinf}
\end{align}
}

\blue{
Furthermore, by the strong law of large numbers, $\hatT_{Y^n}\to P_j$ under hypothesis $\mathrm{H}_j$ and $\hatT_{X_i^N}\to P_i$ for each $i\in[M]$ when $n\to\infty$. Then with the continuity of KL-divergence, we have $\mathrm{GJS}(\hatT_{X_i^N},\hatT_{Y^n},\alpha)\to\mathrm{GJS}(P_i,P_j,\alpha)$ as $n\to\infty$. Recall the definition of the stopping time $\tau_j$, we have
\begin{align}
\label{le}\min\limits_{i\in\calM_j}\mathrm{GJS}\Big(\hatT_{X_i^{N_{\tau_j-1}}},\hatT_{Y^{\tau_j-1}},\alpha\Big)\le g(\beta,\tau_j-1).
\end{align}
When $\tau_j\to\infty$, both $g(\beta,\tau_j)$ and $g(\beta,\tau_j-1)\to-\frac{\log\beta}{\tau_j}$.
Consequently, combining \eqref{ge}, \eqref{tautoinf} and \eqref{le}, we conclude that
\begin{align}\label{prob_conver}
\lim\limits_{\beta\to 0}-\frac{\tau_j}{\log\beta}=\min\limits_{i\in\calM_j}\frac{1}{\mathrm{GJS}(P_i,P_j,\alpha)}.
\end{align}
}
\end{proof}

\blue{
To go from the convergence in probability to convergence in mean, it suffices to prove
that the sequence of random variables $-\frac{\tau_j}{\log\beta}$ is uniformly integrable as $\beta\to 0$. To prove the uniformly integrable, we need the following lemma.
\begin{lemma}\label{tauj_ge}
Given $k\in\bbN,~k\ge 1$, there exists $(n^\prime,c)\in\bbR_+^2$ such that $\bbP_j\{\tau_j\ge k\}\le\frac{1}{\beta}\exp\{-ck\}$ for any $k\ge n^\prime$.
\end{lemma}
}
\begin{proof}
\blue{
Given $\varepsilon\in\bbR_+$, we have
\begin{align}
\nn&\bbP_j\{\tau_j\ge k\}\\*
&\le\bbP_j\Big\{\forall i\in\calM_j,\mathrm{GJS}\Big(\hatT_{X_i^{N_{k-1}}},\hatT_{Y^{k-1}},\alpha\Big)\le g(\beta,k-1)\Big\}\\
&\le\max\limits_{i\in\calM_j}\bbP_j\Big\{\mathrm{GJS}\Big(\hatT_{X_i^{N_{k-1}}},\hatT_{Y^{k-1}},\alpha\Big)\le g(\beta,k-1)\Big\}\\
\nn&\le\max\limits_{i\in\calM_j}\bbP_j\Big\{\mathrm{GJS}\Big(\hatT_{X_i^{N_{k-1}}},\hatT_{Y^{k-1}},\alpha\Big)\le  g(\beta,k-1)~\mathrm{and}~D\Big(\hatT_{X_i^{N_{k-1}}}||P_i\Big)\le\varepsilon~\mathrm{and}~D(\hatT_{Y^{k-1}}||P_j)\le\varepsilon\Big\}\\*
&\quad+\max\limits_{i\in\calM_j}\bbP_j\Big\{D\Big(\hatT_{X_i^{N_{k-1}}}||P_i\Big)>\varepsilon\Big\}
+\bbP_j\big\{D(\hatT_{Y^{k-1}}||P_j)>\varepsilon\big\}.\label{longterm}
\end{align}
}

\blue{
The first term of \eqref{longterm} can be upper bounded as follows:
\begin{align}
\nn&\max\limits_{i\in\calM_j}\bbP_j\Big\{\mathrm{GJS}\Big(\hatT_{X_i^{N_{k-1}}},\hatT_{Y^{k-1}},\alpha\Big)\le  g(\beta,k-1)~\mathrm{and}~D\Big(\hatT_{X_i^{N_{k-1}}}||P_i\Big)\le\varepsilon~\mathrm{and}~D(\hatT_{Y^{k-1}}||P_j)\le\varepsilon\Big\}\\*
\nn&\le\max\limits_{i\in\calM_j}\bbP_j\Big\{\mathrm{GJS}\Big(\hatT_{X_i^{N_{k-1}}},\hatT_{Y^{k-1}},\alpha\Big)\le  g(\beta,k-1)+\mathrm{GJS}\Big(\hatT_{X_j^{N_{k-1}}},\hatT_{Y^{k-1}},\alpha\Big),~\mathrm{and}~D\Big(\hatT_{X_i^{N_{k-1}}}||P_i\Big)\le\varepsilon,\\*
&\qquad\mathrm{and}~D(\hatT_{Y^{k-1}}||P_j)\le\varepsilon\Big\}\label{add}\\
&\le\bbP_j\Big\{\mathrm{GJS}\Big(\hatT_{X_j^{N_{k-1}}},\hatT_{Y^{k-1}},\alpha\Big)\ge \eta-g(\beta,k-1)\Big\}\label{a}\\
\label{similar}&\le\frac{1}{\beta(|\calX|-1)}k^{3|\calX|}(N_{k-1}+1)^{2|\calX|}\exp\{-(k-1)\eta\},
\end{align}
where \eqref{add} follows from that $\mathrm{GJS}\Big(\hatT_{X_j^{N_{k-1}}},\hatT_{Y^{k-1}},\alpha\Big)$ is nonnegative, \eqref{a} follows from that when $\varepsilon$ is chosen to be sufficiently small and $P_i\neq P_j$, if $D(Q_1||P_i)\le\varepsilon$ and $D(Q_2||P_j)\le\varepsilon$, $\exists~\eta>0$ such that $\mathrm{GJS}(Q_1,Q_2,\alpha)>\eta$~\cite[Lemma 3]{Ihwang2022sequential}, \eqref{similar} follows from the similar manner as Eq. \eqref{knumber}.
}

\blue{
Other two terms of \eqref{longterm} can be upper bounded as follows using the upper bound of the probability of a type class:
\begin{align}
\max\limits_{i\in\calM_j}\bbP_j\Big\{D\Big(\hatT_{X_i^{N_{k-1}}}||P_i\Big)>\varepsilon\Big\}
+\bbP_j\big\{D(\hatT_{Y^{k-1}}||P_j)>\varepsilon\big\}
\le (N_{k-1}+1)^{|\calX|}\exp\{-N_{k-1}\varepsilon\}+k^{|\calX|}\exp\{-(k-1)\varepsilon\}\label{epsilon}.
\end{align}
Thus combining \eqref{similar} and \eqref{epsilon}, we have
\begin{align}
\bbP_j\{\tau_j\ge k\}\le\frac{1}{\beta}\exp\{-ck\},\quad\forall k\ge n^\prime~\mathrm{for}~\mathrm{some}~c~\mathrm{and}~n^\prime>0.
\end{align}
}
\end{proof}

\blue{
Using Lemma \ref{tauj_ge} and \cite[Lemma 5]{Ihwang2022sequential}, we have that $\{-\frac{\tau_j}{\log\beta}\}_{\beta\in(0,0.9]}$ is uniformly integrable.
Therefore, we can obtain the convergence in mean of $\{-\frac{\tau_j}{\log\beta}\}_{\beta\in(0,0.9]}$ from the convergence in probability in Eq. \eqref{prob_conver}.
Furthermore, we have $\tau_j\le\tau_{\rm{seq}}$ by definition in \eqref{def:tauj}. Then we have
\begin{align}
E_j(\Phi_{\rm{seq}}^{(M)}|\bP)=\liminf_{\beta\to 0}\frac{-\log\beta}{\mathbb{E}_j[\tau_{\rm{seq}}]}\ge\liminf_{\beta\to 0}\frac{-\log\beta}{\mathbb{E}_j[\tau_j]}=\min\limits_{i\in\calM_j}\mathrm{GJS}(P_i,P_j,\alpha).
\end{align}
}

\subsection{Converse Proof of Theorem \ref{wang:mary}}\label{Mseqcon}
\blue{
In the following, we drive the converse result for the optimal sequential test under error probability universality constraint with $\beta$ when $\beta\to 0$.
}

\blue{
Given $(p,q)\in(0,1)^2$, define the binary KL-divergence as follows.
\begin{align}
d(p,q):=p\log\frac{p}{q}+(1-p)\log\frac{1-p}{1-q}.
\end{align}
The first-order derivative of $d(p,q)$ on $q$ is as follows:
\begin{align}
\frac{\partial d(p,q)}{\partial q}=\frac{q-p}{q(1-q)}.
\end{align}
Thus, $d(p,q)$ is increasing in $q$ when $q>p$ and decreasing in $q$ when $p>q$.
}

\blue{
Given $j\in[M]$ and any $i\in\calM_j$, define the event $\calW:=\{\phi^{(M)}_\tau(\bX^N,Y^n)=i\}$. For any two tuple of distributions $\bP=\{P_t\}_{t=1}^{M}\in\calP(\calX)^M$ and $\tilde\bP=\{\tilP_t\}_{t=1}^{M}\in\calP(\calX)^M$, and any sequential test $\Phi^{(M)}=(\tau,\phi^{(M)}_\tau)$, following~\cite[Lemma 1]{Ihwang2022sequential}, we have
\begin{align}
d\big(\bbP_i(\calW),\tilde\bbP_j(\calW)\big)
\le\mathbb{E}_i[\tau]D(P_i||\tilP_j)+\mathbb{E}_i[N_\tau]\big(D(P_i||\tilP_i)+D(P_j||\tilP_j)\big).\label{ED}
\end{align}
}

\blue{
Recall the definition of error probability universality constraint with $\beta$ in Eq. \eqref{sequential_universal}. Given $j\in[M]$ and any $i\in\calM_j$, when $\beta\to 0$, we have
\begin{align}
\bbP_i(\calW)&=\bbP_i\big(\phi^{(M)}_\tau(\bX^N,Y^n)=i\big)=1-\beta_i(\Phi^{(M)}|\bP)\to 1,\\*
\tilde\bbP_j(\calW)&=\tilde\bbP_j\big(\phi^{(M)}_\tau(\bX^N,Y^n)=i\big)\le\tilde\beta_j(\Phi^{(M)}|\tilde\bP)\to 0.
\end{align}
Thus we obtain that $\bbP_i(\calW)>\tilde\bbP_j(\calW)$ and $1-\beta_i(\Phi^{(M)}|\bP)>\tilde\beta_j(\Phi^{(M)}|\tilde\bP)$. Furthermore, we have
\begin{align}
d\big(\bbP_i(\calW),\tilde\bbP_j(\calW)\big)&\ge d\big(1-\beta_i(\Phi^{(M)}|\bP),\tilde\beta_j(\Phi^{(M)}|\tilde\bP)\big)\label{d_func}\\*
&=-\log\tilde\beta_j(\Phi^{(M)}|\tilde\bP)\label{huslemma2}\\
&\ge-\log\beta,\label{betage}
\end{align}
where \eqref{d_func} follows from that $d(p,q)$ is decreasing in $q$ when $p>q$, \eqref{huslemma2} follows from \cite[Lemma 2]{Ihwang2022sequential} when $\beta\to 0$ and \eqref{betage} follows from the definition of error probability universality constraint.
}

\blue{
Given $j\in[M]$, combining \eqref{ED} and \eqref{betage}, we have
\begin{align}
-\log\beta\le\min_{i\in\calM_j}\mathbb{E}_i[\tau]D(P_i||\tilP_j)+\mathbb{E}_i[N_\tau]\big(D(P_i||\tilP_i)+D(P_j||\tilP_j)\big).
\end{align}
It is clear that $\mathbb{E}_i[\tau]\to\infty$ when $\beta\to 0$.
For any sequential test $\Phi^{(M)}=(\tau,\phi)$ satisfying the error probability universality constraint with $\beta$ and any tuple of distributions $\tilde\bP\in\calP(\calX)^M$, the type-$j$ error exponent satisfies
\begin{align}
\label{con:ej}\tilE_j(\Phi^{(M)}|\tilde\bP)\le\min_{i\in\calM_j}\Big(D(P_i||\tilP_j)+\alpha\big(D(P_i||\tilP_i)+D(P_j||\tilP_j)\big)\Big).
\end{align}
Since \eqref{con:ej} is true for all $\tilde\bP\in\calP(\calX)^M$, we can minimize the RHS of \eqref{con:ej} with respect to $\tilde\bP$ and obtain
\begin{align}
\tilE_j(\Phi^{(M)}|\tilde\bP)\le\min_{i\in\calM_j}\mathrm{GJS}(\tilP_i,\tilP_j,\alpha).
\end{align}
}

\subsection{Achievability Proof of Theorem \ref{M_exponents}}\label{Maryach}

Given any $\bP\in\calP(\calX)^M$, for each $j\in[M]$, the achievable type-$j$ error probability satisfies
\begin{align}
\beta_j\big(\Phi_{\rm{tp}}^{(M)}|\bP\big)=\bbP_j\big\{\tau=n,\phi_n^{(M)}(\bX,Y^n)\neq\rmH_j\big\}+\bbP_j\big\{\tau=kn,\phi_{kn}^{(M)}(\bX,Y^{kn})\neq\rmH_j\big\}\label{error:Mclassify}.
\end{align}

The first term of \eqref{error:Mclassify}, i.e., the type-$j$ error probability of the first phase, can be upper bounded as follows:
\begin{align}
\nn&\bbP_j\big\{\tau=n,\phi_n^{(M)}(\bX,Y^n)\neq \rmH_j\big\}\\*
&=\bbP_j\Big\{i^*(\bX^N,Y^n)\neq j,~\mathrm{GJS}\Big(\hatT_{X_k^N},\hatT_{Y^n},\alpha\Big)\ge\lambda_k,\forall k\neq i^*(\bX^N,Y^n)\Big\}\\
&\le\bbP_j\Big\{\mathrm{GJS}\Big(\hatT_{X_j^N},\hatT_{Y^n},\alpha\Big)\ge\lambda_j\Big\}\\
\label{Mary:phase1}&\le\exp\big\{-n\lambda_j+|\calX|\log(N+n+1)\big\},
\end{align}
where \eqref{Mary:phase1} follows from the similar manner as Eq. \eqref{knumber}.

The second term of \eqref{error:Mclassify}, i.e., the type-$j$ error probability of the second phase, can be decomposed as the excess-length probability $\bbP_j\{\tau=kn\}$ and the error probability $\bbP_j\{\phi_{kn}(X_1^N,X_2^N,Y^{kn})\neq \rmH_j\}$.

Recall the definition of $\calM_{\mathrm{dis}}$ in Eq. \eqref{Mdis}. Given $(i,l)\in\calM_{\mathrm{dis}}$ and $(\lambda_1,\ldots,\lambda_M)\in\bbR_+^M$, define a set $\calB:=\{\bQ\in\calP^N(\calX)\times\calP^N(\calX)\times\calP^n(\calX):\mathrm{GJS}(Q_1,Q_3,\alpha)\le\lambda_l,\mathrm{GJS}(Q_2,Q_3,\alpha)\le\lambda_i\}$. Under hypothesis $\rmH_j$, the excess-length probability, i.e., $\bbP_j\{\tau=kn\}$, can be upper bounded as follows:
\begin{align}
\nn&\bbP_j\{\tau=kn\}\\*
&=\bbP_j\Big\{\exists(i,l)\in\calM_{\mathrm{dis}},~\mathrm{s.t.}~\mathrm{GJS}\Big(\hatT_{X_l^N},\hatT_{Y^n},\alpha\Big)<\lambda_l~\mathrm{and}~\mathrm{GJS}\Big(\hatT_{X_i^N},\hatT_{Y^n},\alpha\Big)<\lambda_i\Big\}\\
&\le\sum\limits_{(i,l)\in\calM_{\mathrm{dis}}}\bbP_j\Big\{\mathrm{GJS}\Big(\hatT_{X_l^N},\hatT_{Y^n},\alpha\Big)<\lambda_l,\mathrm{GJS}\Big(\hatT_{X_i^N},\hatT_{Y^n},\alpha\Big)<\lambda_i\Big\}\\
&=\sum\limits_{(i,l)\in\calM_{\mathrm{dis}}}\sum\limits_{x_l^N,x_i^N,y^n:\mathrm{GJS}\big(\hatT_{x_l^N},\hatT_{y^n},\alpha
\big)\le\lambda_l, \mathrm{GJS}\big(\hatT_{x_i^N},\hatT_{y^n},\alpha
\big)\le\lambda_i} P_l(x_l^N)P_i(x_i^N)P_j(y^n) \\
&=\sum\limits_{(i,l)\in\calM_{\mathrm{dis}}}\sum\limits_{\bQ\in\calB} P_l^N(\calT_{Q_1}^N)P_i^N(\calT_{Q_2}^N)P_j^n(\calT_{Q_3}^n) \\
&\le\sum\limits_{(i,l)\in\calM_{\mathrm{dis}}}\sum\limits_{\bQ\in\calB} \exp\big\{-n D(\alpha,\bQ|P_i,P_l,P_j)\big\}\label{D_upper}\\
&\le M(M-1)\max\limits_{(i,l)\in\calM_{\mathrm{dis}}}\sum\limits_{\bQ\in\calB} \exp\big\{-n D(\alpha,\bQ|P_i,P_l,P_j)\big\}\\
&\le M(M-1)|\calP^N(\calX)||\calP^N(\calX)||\calP^n(\calX)|\max\limits_{(i,l)\in\calM_{\mathrm{dis}}}\max\limits_{\bQ\in\calB}\exp\big\{-nD(\alpha,\bQ|P_i,P_l,P_j)\big\}\\
&\le M(M-1)|\calP^N(\calX)||\calP^N(\calX)||\calP^n(\calX)|\exp\big\{-n F_j(\alpha,\lambda^M|\bP)\big\}\label{Fj}\\
&\le\exp\big\{-nF_j(\alpha,\lambda^M|\bP)+|\calX|\log(N+1)^2(n+1)+\log M(M-1)\big\}\label{fnumber}.
\end{align}
where \eqref{D_upper} follows from the upper bound of the probability of a type class, \eqref{Fj} follows from the definition of $F_j(\alpha,\lambda^M|\bP)$ in Eq. \eqref{def:Fj} and \eqref{fnumber} follows from the fact that the number of the set of types of length $n$ satisfies $|\calP^n(\calX)|\le(n+1)^{|\calX|}$. Thus, for each $j\in[M]$ and $\lambda^M=(\lambda_1,\ldots,\lambda_M)\in \tilG(\gamma)$, we have
\begin{align}\label{Mary:gamma}
\mathop{\lim\inf}\limits_{n\to\infty}-\frac{1}{n}\log\bbP_j\{\tau=kn\}\ge F_j(\alpha,\lambda^M|\bP)\ge\gamma.
\end{align}
where $\tilG(\gamma)=\{ \bar{\lambda}^M\in\bbR_+^M:~\min\limits_{j\in[M]}F_j(\alpha,\bar\lambda^M|\bP)\ge\gamma\}$.

Define $\calS:=\{\bQ\in\calP^N(\calX)\times\calP^N(\calX)\times\calP^{kn}(\calX):\mathrm{GJS}(Q_1,Q_3,\alpha)<\mathrm{GJS}(Q_2,Q_3,\alpha)\}$. Recall the definition of $\calM_j=\{i\in[M]:i\neq j\}$. Under hypothesis $\rmH_j$, the error probability, i.e., $\bbP_j\{\phi_{kn}(X_1^N,X_2^N,Y^{kn})\neq \rmH_j\}$, is upper bounded as follows:
\begin{align}
\nn&\bbP_j\big\{\phi_{kn}^{(M)}(\bX^N,Y^{kn})\neq\rmH_j\big\}\\*
&=\bbP_j\{i^*(\bX^N,Y^{kn})\neq j\}\\
&=\bbP_j\Big\{\exists i\in\calM_j,~\mathrm{s.t.}~\mathrm{GJS}\Big(\hatT_{X_i^N},\hatT_{Y^{kn}},\alpha\!\Big)<\mathrm{GJS}\Big(\hatT_{X_j^N},\hatT_{Y^{kn}},\alpha\Big)\Big\}\\
&\le\sum\limits_{i\in\calM_j}\bbP_j\Big\{\mathrm{GJS}\Big(\hatT_{X_i^N},\hatT_{Y^{kn}},\alpha\Big)<\mathrm{GJS}\Big(\hatT_{X_j^N},\hatT_{Y^{kn}},\alpha\Big)\Big\}\\
&=\sum\limits_{i\in\calM_j}\sum\limits_{x_i^N,x_j^N,y^{kn}:\mathrm{GJS}\big(\hatT_{x_i^N},\hatT_{y^{kn}},\alpha
\big)<\mathrm{GJS}\big(\hatT_{x_j^N},\hatT_{y^{kn}},\alpha
\big)} P_i(x_i^N)P_j(x_j^N)P_j(y^{kn}) \\
&=\sum\limits_{i\in\calM_j}\sum\limits_{\bQ\in\calS}P_i^N(\calT_{Q_1}^N)P_j^N(\calT_{Q_2}^N)P_j^n(\calT_{Q_3}^{kn})\\
&\le\sum\limits_{i\in\calM_j}\sum\limits_{\bQ\in\calS}\exp\big\{-nU(k,\alpha,\bQ|P_i,P_j)\big\}\\
\label{Lj}&\le\exp\big\{-nL_j(k,\alpha|\bP)+|\calX|\log(N+1)^2(n+1)+\log (M-1)\big\}.
\end{align}
where \eqref{Lj} follows from the definition of $L_j(k,\alpha|\bP)$ in Eq. \eqref{def:Lj}.

When $\gamma>0$, we have $\mathbb{E}[\tau]=n+kn\cdot\exp\{-n\gamma\}$, which satisfies that $\lim_{n\to\infty}\frac{\mathbb{E}[\tau]}{n}=1$. Combining the results in Eq. \eqref{Mary:phase1}, Eq. \eqref{Mary:gamma}, and Eq. \eqref{Lj}, for any $\lambda^M\in \tilG(\gamma)$, we have
\begin{align}
E_j(\Phi_{\rm{tp}}^{(M)}|\bP)\ge\min\big\{\lambda_j,L_j(k,\alpha|\bP)+\gamma\big\},
\end{align}
Then the Bayesian error exponent of the two-phase test for multiple classification satisfies
\begin{align}\label{mary:bayesian}
E_\mathrm{Bayesian}\big(\Phi_{\mathrm{tp}}^{(M)}|\bP\big)\ge\max\limits_{\lambda^M\in\tilG(\gamma)}\min\limits_{j\in[M]}\min\big\{\lambda_j,L_j(k,\alpha|\bP)+\gamma\}.
\end{align}

\section{Conclusion}
\blue{
We proposed a two-phase test, analyzed its achievable error exponents for $M$-ary classification and specialized the results to binary classification when $M=2$ and to $M$-ary hypothesis testing when $\alpha\to\infty$. To provide theoretical optimality guarantee for our test, we derived the achievable error exponent of an optimal sequential test under the error probability universality constraint. We showed that with proper test design parameters, our two-phase tests achieved the performance close to the optimal sequential test with the simple construction of a fixed-length test. Our results significantly generalized the results of Lalitha and Javidi~\cite{lalithaalmost} for binary hypothesis testing by allowing multiple decision outcomes and considering the more practical scenario where the generating distribution under each hypothesis is unknown but available from training sequences. Furthermore, we discussed the influence of the ratio of the lengths of training sequences and testing sequence on the performance of our tests and illustrated our results with numerical examples.
}

There are several avenues for future research. Firstly, we only considered discrete alphabet sequences in this paper. It is of interest to generalize our results to continuous sequences. In this case, the method of types failed and novel ideas such as kernel methods might be required~\cite{zou2017nonparametric}. Secondly, we focused on hypothesis testing and statistical classification in this paper. We believe that two-phase tests would also be applicable to other statistical inference problems, e.g., distributed detection~\cite{he2019distributed,tenney1981detection}, quickest change-point detection~\cite{he2021optimal,pettitt1979non} and outlier hypothesis testing~\cite{li2014,li2017universal,zhou2022second}.

\blue{
\section*{Acknowledgments}
The authors acknowledge two anonymous reviewers of the previous version of the manuscript for providing many helpful comments and suggestions, which significantly improve the quality of the current manuscript.
}

\bibliographystyle{IEEEtran}
\bibliography{IEEEfull_paper}
\end{document}